# Complex geometry and pre-metric electromagnetism

D.H. Delphenich[†]
Physics Department
Bethany College
Lindsborg, KS 67456

*Abstract.* The intimate link between complex geometry and the problem of the pre-metric formulation of electromagnetism is examined. In particular, the relationship between 3+1 decompositions of $\mathbb{R}^4$ and decompositions of $\Lambda_2(\mathbb{R}^4)$ into real and imaginary subspaces relative to a choice of complex structure is emphasized. The role of the various scalar products on $\Lambda_2(\mathbb{R}^4)$ that are defined in terms of a volume element on $\mathbb{R}^4$ and a complex structure on $\Lambda_2(\mathbb{R}^4)$ that makes it $\mathbb{C}$-linear isomorphic to $\mathbb{C}^3$ is discussed in the context of formulating a theory of electromagnetism in which the Lorentzian metric on spacetime follows as a consequence of the existence of electromagnetic waves, not a prior assumption.

Contents

---

[†] E-mail: delphenichd@bethanylb.edu .

**0. Introduction.** Since the early days of Einstein's theory of gravitation, it had been suspected by some researchers, such as Van Dantzig [**1**], Kottler [**2**], and Cartan [**3**], that the introduction of a metric into the geometrical structure of the spacetime manifold, although fundamental to the theory of the gravitational interaction, was not only mathematically unnecessary in the theory of the electromagnetic interaction, but from a physical standpoint it was also naïve. This is because the metric structure of spacetime is so intimately related to the electromagnetic structure by way of the propagation of electromagnetic waves that one rather suspects that the metric structure might plausibly follow as a consequence of the electromagnetic structure, rather than represent a basic component of that structure. Indeed, it is interesting that the origin of Einstein's theory of gravitation – viz., general relativity – was in his study of the electrodynamics of moving particles. In particular, the Lorentzian metric that was later generalized to embody the essence of spacetime geometry first appeared as the symbol of the d'Alembertian operator that governed the propagation of electromagnetic waves. This also has the effect of saying that gravitation is not entirely independent of electromagnetism since gravitational waves are governed by the same light cones that are defined by the demands of electromagnetic wave propagation.

Since the only place in which the spacetime metric $g$ intervenes in Maxwell's equations for electromagnetism is in the definition of the Hodge * operator that takes the Minkowski field strength 2-form $F$ to its Hodge dual 2-form $*F$, a first attempt at eliminating $g$ from the equations would be to introduce the duality isomorphism for 2-forms as the fundamental object in place of $g$. This is effectively equivalent to going from metric geometry to conformal geometry, since such a choice of * is known to be equivalent to a conformal class of metrics.

However, physical considerations, in the form of the more general formulation of Maxwell's equations for electromagnetic fields in continuous media, suggest that one should treat the electromagnetic field strength 2-form $F$ and the electromagnetic induction 2-vector field $\mathfrak{H}$ as distinct fields that are related to each other by an electromagnetic constitutive law for the medium. In general this constitutive law takes the form of a fiber-preserving diffeomorphism that covers the identity and maps the space of 2-forms $\Lambda^2(M)$ on the spacetime manifold $M$ onto the space of 2-vector fields $\Lambda_2(M)$. Now, the constitutive law for the classical spacetime electromagnetic vacuum is usually modeled as simply a multiplication by a constant, in the form of $4\pi/c_0 = 4\pi\sqrt{\varepsilon_0\mu_0}$, times the isomorphism $\iota \wedge \iota$: $\Lambda^2(M) \to \Lambda_2(M)$ that one derives from the metric isomorphism $\iota$: $T^*(M) \to T(M)$. Hence, one might generalize from this by postulating a linear constitutive law – i.e., a linear fiber-preserving isomorphism that covers the identity $\chi$: $\Lambda^2(M) \to \Lambda_2(M)$, $F \mapsto \chi(F) = \mathfrak{H}$.

Of course, this linearity may well be exactly the limiting approximation that defines the transition from Maxwell's linear equations of electromagnetism into the world of electrodynamics at the atomic to subatomic level, which seems to depend crucially upon the existence of vacuum polarization at high field strengths, a process that would demand a nonlinear constitutive law. Nevertheless, it is still an illustrative first step in that direction to examine the geometrical nature of the linear constitutive laws.

If one starts with the linear constitutive law in the above form then $\chi$ is also equivalent to a scalar product on the vector bundle $\Lambda^2(M)$, namely, $<\alpha, \beta> = \chi(\alpha, \beta) = \alpha(\chi(\beta))$, as long as one also assumes the symmetry of $\chi$, which amounts to the physical condition called electric/magnetic reciprocity. Our ultimate objective is then to see if we can relate this scalar product on $\Lambda^2(M)$ to an induced scalar product on $T(M)$ of Lorentzian type, or at least a conformal class of Lorentzian metrics.

One way of resolving this when $M$ is four-dimensional, orientable, oriented, and given a choice of unit-volume element $\mathcal{V}$ is to compose the linear isomorphism $\chi$ with the Poincaré duality isomorphism #: $\Lambda_2(M) \to \Lambda^2(M)$ that is defined by $\mathcal{V}$ to get an isomorphism of $\Lambda^2(M)$ with itself that we also call * ([1]), although we do not have this kind of Hodge duality defined for any other $k$-forms than 2-forms. One the sees that there are two basic types of such an isomorphism depending on which sign one chooses in the equation $*^2 = \pm I$. One then finds that the constitutive law $\chi$ makes $*^2 = -I$ iff there is a conformal class of Lorentzian metrics on $T(M)$ that makes * the resulting Hodge *. Ultimately, one also needs to physically account for the reduction from a conformal class to a choice of metric.

The fact that $*^2 = -I$ implies that * defines an almost complex structure on $\Lambda^2(M)$, along with a decomposition of $\Lambda^2(M) = \Lambda^2_{Re}(M) \oplus \Lambda^2_{Im}(M)$, which is not, however, unique. Hence, it is convenient to regard the fibers of $\Lambda^2(M)$ as behaving like $\mathbb{C}^3$, so the complex projectivization of $\Lambda^2(M)$ has fibers like $\mathbb{CP}^2$, which is also a compact four-dimensional real manifold. Of course, one has to establish the action of $\mathbb{C}$ on $\Lambda^2(M)$ in order to make this reduction to $\mathbb{CP}^2$ natural. However, since one already has an action of $\mathbb{R}$ by scalar multiplication, all that remains to be done is to define an action of $U(1)$, which follows naturally from the introduction of * by way of the "duality" rotation that continuously rotates real 2-forms into imaginary ones and vice versa.

So far, we replaced a Lorentzian metric on $T(M)$ with a volume element $\mathcal{V}$ on $T(M)$ and an almost complex structure * on $\Lambda^2(M)$. The problem then posed to theoretical physics is that of gaining more physical intuition concerning the nature of these otherwise purely mathematical constructions. At first, this seems unlikely, since these structures might seem overly mathematically general in scope, at least to most physicists. However, this is simply due to the fact that for most theoretical physicists "geometry" means *metric* geometry, which seems to mostly relate to gravitation. The most that mainstream physicists are willing to generalize beyond metric geometry at present seems to be conformal geometry, which describes the motion of massless matter, which is represented by some, but not all, of the gauge particles at the elementary level ([2]).

A more general conception of geometry that was widely accepted in the earlier part of the Twentieth Century was the view that the most general geometry is projective geometry, which then reduces, in turn, to affine geometry, conformal geometry, and finally metric geometry. The conceptual mechanism for this reduction is first, the choice of a point in projective affine space and an affine subspace that does not contain that point and generates the space of lines through the chosen point, then a choice of angle measure on that space of lines, and finally, a choice of metric on the chosen affine subspace. In a sense, our shift in emphasis will be from the geometry of Riemann to the geometry of Grassmann, Plücker, Cayley, and Klein.

---

[1] Strictly speaking, the composed isomorphism $\# \circ \chi$ is only equal to * up to a multiplicative factor, as we shall discuss in due course.

[2] Of course, we are treating the experimental proof of the non-zero mass of neutrinos as a *fait accompli.*

One finds that the otherwise esoteric structures that we introduced – in the form of $\mathcal{V}$ and * – are not only rife with deeper mathematical consequences in the context of projective geometry – in particular, complex projective geometry – but also deeper physical significance, when one first re-examines the Minkowski geometry of special relativity in the context of projective geometry, and then examines the geometrical setting for pre-metric electromagnetism.

This article has the more restricted intent of simply laying the mathematical and physical foundations for further discussion of the role of projective geometry by first establishing the notion that the existence of a $\mathbb{C}$-linear isomorphism of the space $\Lambda_2(\mathbb{R}^4)$ of bivectors on $\mathbb{R}^4$ implies that the most immediately relevant types of projective geometry than one should examine the physical consequences of are the real projective geometry of lines and 3-planes in $\mathbb{R}^4$, i.e., the spaces $\mathbb{R}P^3$ and $\mathbb{R}P^{3*}$, as well as the complex projective geometry of $\mathbb{C}P^2$, which can be viewed as the space of all complex lines – or "duality 2-planes" – in $\mathbb{C}^3$ or $\Lambda_2(\mathbb{R}^4)$, respectively. The actual discussion of the projective geometric aspects of the aforementioned isomorphism will follow in a subsequent study that is still in preparation.

Mention will also be made of the more established methods of complex geometry by which the space $\Lambda_2(\mathbb{R}^4)$ is regarded as $\mathbb{C}$-linear isomorphic to the subspace of the complexification $\Lambda_2(\mathbb{R}^4) \otimes \mathbb{C}$ that consists of self-dual complex bivectors. However, the reason that the present article will not start with that fact is simply that when $\Lambda_2(\mathbb{R}^4)$ is given a complex structure to begin with, it becomes somewhat redundant to complexify the space as well, only to use a subspace of the resulting complexified space, but not the whole space.

**1. The geometry of** $\Lambda_2(\mathbb{R}^4)$. We first pay particular attention to the special nature of the second exterior product $\Lambda_2(\mathbb{R}^4)$ over a four-dimensional real vector space; we use $\mathbb{R}^4$ simply for the sake of specificity. This is because, although our choice of frame for $\mathbb{R}^4$ will usually be general, nevertheless $\mathbb{R}^4$ gives one the option of choosing a canonical frame, namely, the one whose column vectors collectively define the 4×4 identity matrix.

We first point out a subtlety that might possibly be a source of later confusion: although $\Lambda_2(\mathbb{R}^4)$ is *spanned* by all finite sums of exterior products of vectors in $\mathbb{R}^4$, nevertheless, just as a vector in $\mathbb{R}^4$ can be represented in an infinitude of ways as a finite sum of other such vectors, similarly, any element of $\Lambda_2(\mathbb{R}^4)$ can be represented in an infinitude of ways as a finite sum of exterior products; i.e., an element of $\Lambda_2(\mathbb{R}^4)$ is an *equivalence class* of all such finite sums. It is this ambiguity of expression that makes the use of frames in one's logic at times tenuous, since a bivector can be a single exterior product with respect to one frame and a sum of exterior products with respect to another. Hence, when we speak of two bivectors having a common exterior factor, we really mean that there is at least one pair of expressions for each that has this property. Otherwise, we could not always see that $\mathbf{a} \wedge \mathbf{b}$ and $\frac{1}{2} a^\mu b^\nu \mathbf{e}_\mu \wedge \mathbf{e}_\nu$ had the common exterior factors $\mathbf{a}$ and $\mathbf{b}$, which should be otherwise obvious.

Now, since any fraction over the natural numbers can be reduced to a simplest form that represents a unique rational number, one then wonders if a similar situation applies to bivectors. Sadly, one can only go so far in the direction of finding a "canonical" form for the expression of a given bivector by a finite sum of exterior products, but, as we shall now see, one can still speak of a smaller equivalence class when one introduces the notion of the "rank" of a bivector.

*a. The rank of a bivector.* If $\alpha \in \Lambda_2(\mathbb{R}^4)$ then its *rank* is defined by the equivalent conditions:

- *i*) The minimum integer $r$ such that the $r^{th}$ exterior power of $\alpha$ is zero.
- *ii*) The minimum number of non-zero vectors whose exterior products can add up to $\alpha$.
- *iii*) The number $r$ such that the space of all $\mathbf{x} \in \mathbb{R}^4$ such that $\mathbf{x} \wedge \alpha = 0$ is $4-r$-dimensional.
- *iv*) The rank of the component matrix of $\alpha$ with respect to any frame on $\mathbb{R}^4$.

From the first condition, it is clear that the rank of any bivector on $\mathbb{R}^4$ can be 0, 2 or 4. The case of rank zero is simply the trivial case of $\alpha = 0$ and will generally be ignored.

From the second condition, we see that a bivector $\alpha$ of rank two is expressible in the form $\mathbf{a} \wedge \mathbf{b}$ where $\mathbf{a}$ and $\mathbf{b}$ are linearly dependent; i.e., it is *decomposable* or, as it is sometimes phrased, *simple.* Furthermore, the pair of vectors $\mathbf{a}$ and $\mathbf{b}$ define a 2-frame in $\mathbb{R}^4$ that spans a 2-plane $\Pi_2$. If $\mathbf{a}'$ and $\mathbf{b}'$ are two other linearly independent vectors in $\Pi_2$ then there is an invertible linear relation between the two 2-frames $\mathbf{a}' = a\mathbf{a} + b\mathbf{b}$, $\mathbf{b}' = c\mathbf{a} + d\mathbf{b}$. Hence, $\mathbf{a}' \wedge \mathbf{b}' = (ac - bd)\, \mathbf{a} \wedge \mathbf{b}$, so any pair of linearly independent vectors that span $\Pi_2$ will produce bivectors that generate the same (real) line $[\alpha]$ in $\Lambda_2(\mathbb{R}^4)$; if the invertible 2×2 matrix that relates the two 2-frames has unit determinant − i.e., is an element of $SL(2; \mathbb{R})$ – then $\mathbf{a}' \wedge \mathbf{b}' = \mathbf{a} \wedge \mathbf{b}$. From this, we can see that the decomposition of $\alpha$ into a product of vectors is not unique, but the association of $\alpha$ with a 2-plane *is* unique. Indeed, one can see that a decomposable bivector is really an equivalence class of other such bivectors, namely, the orbit of any one of them under the action of $SL(2; \mathbb{R})$. As we shall see, the set of all rank-two bivectors on $\mathbb{R}^4$ is a real quadric in $\Lambda_2(\mathbb{R}^4)$ – viz., the *Klein quadric,* – which has dimension five.

If $\alpha$ has rank four then it is expressible in the form $\mathbf{a} \wedge \mathbf{b} + \mathbf{c} \wedge \mathbf{d}$, where $\mathbf{a}, \mathbf{b}, \mathbf{c}, \mathbf{d}$ are linearly independent vectors. Hence, a bivector of rank four is associated with a 4-frame in $\mathbb{R}^4$, but not uniquely, since the same argument as in the rank-two case applies to the terms $\mathbf{a} \wedge \mathbf{b}$ and $\mathbf{c} \wedge \mathbf{d}$. However, since the 2-planes that they span are unique, one can uniquely associate a rank-four bivector with a unique complementary pair of 2-planes in $\mathbb{R}^4$.

*b. Complex structures on $\Lambda_2(\mathbb{R}^4)$.* A *complex structure* on the vector space $\Lambda_2(\mathbb{R}^4)$ is a linear isomorphism of $\Lambda_2(\mathbb{R}^4)$ with itself, which we will denote by *, that satisfies the condition that:

$$*^2 = -I. \tag{1.1}$$

Hence, since $\Lambda_2(\mathbb{R}^4)$ is $\mathbb{R}$-linearly isomorphic to $\mathbb{R}^6$ by any choice of basis and $\mathbb{R}^6 = \mathbb{R}^3 \times \mathbb{R}^3$ is $\mathbb{R}$-linearly isomorphic to $\mathbb{C}^3$, and * is analogous to multiplying elements $\mathbb{C}^3$ by the imaginary scalar $i$, we can show that $\Lambda_2(\mathbb{R}^4)$ is $\mathbb{C}$-linear isomorphic to $\mathbb{C}^3$ if we can define an action of $\mathbb{C}$ on $\Lambda_2(\mathbb{R}^4)$ by scalar multiplication. It is important to note that this isomorphism will not be canonical, though, and will generally depend upon a choice of frame for $\Lambda_2(\mathbb{R}^4)$. However, before we define the multiplication of bivectors by complex scalars, we need to discuss the decomposition of $\Lambda_2(\mathbb{R}^4)$ into "real" and "imaginary" subspaces.

The essential character of any decomposition of $\Lambda_2(\mathbb{R}^4)$ into complementary three-dimensional subspaces: $\Lambda_2(\mathbb{R}^4) = \Lambda_{\mathrm{Re}}(\mathbb{R}^4) \oplus \Lambda_{\mathrm{Im}}(\mathbb{R}^4)$, in which "Re" suggests the "real" bivectors and "Im" suggests "imaginary" ones, is that one must have:

$$*\Lambda_{\mathrm{Re}}(\mathbb{R}^4) = \Lambda_{\mathrm{Im}}(\mathbb{R}^4), \qquad *\Lambda_{\mathrm{Im}}(\mathbb{R}^4) = -\Lambda_{\mathrm{Re}}(\mathbb{R}^4). \tag{1.2}$$

This is necessary in order to make the eventual $\mathbb{C}$-linear isomorphism of $\Lambda_2(\mathbb{R}^4)$ with $\mathbb{C}^3$ consistent with multiplication by * or $i$, respectively.

Since (1.1) implies that the eigenvalues of * are $\pm i$, there are no *real* bivectors that are eigenvectors of *. Hence, we cannot simply use an eigenspace decomposition relative to * for our definition of real and imaginary; similarly, the notion of principal frames is undefined. We shall, nevertheless, discuss the notions of self-duality and anti-self-duality of *complex* bivectors shortly, since they are so well established in the literature of relativity that their omission from the discussion would seem suspicious.

Although we shall define complex scalar multiplication on $\Lambda_2(\mathbb{R}^4)$ and a $\mathbb{C}$-linear isomorphism of $\Lambda_2(\mathbb{R}^4)$ with $\mathbb{C}^3$, so the *complex* eigen-bivectors of * can be said to exist in $\Lambda_2(\mathbb{R}^4)$, nevertheless, we shall find that since the multiplication by $i$ coincides with the * isomorphism the eigen-bivector equation $*\boldsymbol{\alpha} = \pm i\,\boldsymbol{\alpha}$ essentially reduces to an identity.

One can define a canonical decomposition of $\Lambda_2(\mathbb{R}^4)$ into $\Lambda_2^-(\mathbb{R}^4) \oplus \Lambda_2^+(\mathbb{R}^4)$ by polarization using the * isomorphism, namely, $\boldsymbol{\alpha} = \boldsymbol{\alpha}_+ + \boldsymbol{\alpha}_-$, where:

$$\boldsymbol{\alpha}_\pm = \tfrac{1}{2}(\boldsymbol{\alpha} \pm *\boldsymbol{\alpha}). \tag{1.3}$$

One observes that under the * isomorphism:

$$*\Lambda_2^-(\mathbb{R}^4) = \Lambda_2^+(\mathbb{R}^4), \quad *\Lambda_2^+(\mathbb{R}^4) = -\Lambda_2^-(\mathbb{R}^4), \tag{1.4}$$

so this decomposition of $\Lambda_2(\mathbb{R}^4)$ certainly satisfies (1.2). However, although there is a canonical character to this decomposition, and, as we shall discuss shortly, its complexification has been the subject of considerable interest in the theory of gravitation, nevertheless, we shall not be as concerned with this decomposition in the present article. This is because if one examines the analogous polarization in $\mathbb{C}^3$ – namely, $\tfrac{1}{2}(z^i \pm iz^i)$ – then one sees that one does not actually obtain the real and imaginary parts of $z^i$, which come from polarizing $z^i$ using the complex conjugation operator. However, in the real context of $\Lambda_2(\mathbb{R}^4)$, one can define an analogue of complex conjugation only after first defining a decomposition of $\Lambda_2(\mathbb{R}^4)$ into "real" and "imaginary" subspaces. Hence, we shall look elsewhere for our definition of real and imaginary subspaces.

The first thing to emphasize is that such a decomposition is not unique. Indeed, one can start with any 3-plane $\Pi_3$ in $\Lambda_2(\mathbb{R}^4)$ such that $*\Pi_3 \cap \Pi_3 = 0$ as one's $\Lambda_{\mathrm{Re}}(\mathbb{R}^4)$ and define $\Lambda_{Im}(\mathbb{R}^4)$ to be $*\Pi_3$. (We shall return to the contrary case after we have discussed duality rotations.) The arbitrariness of the choice of $\Pi_3$ suggests that there might be a fundamental role played by the subgroup of $GL(6; \mathbb{R})$ that maps 3-planes to other 3-planes while preserving the duality relation. In fact, since any invertible linear map of $\Lambda_2(\mathbb{R}^4)$ to itself will take a 3-plane to another 3-plane, the only restriction is imposed by the fact that not all invertible linear maps preserve duality. As we shall discuss in the next section, when this is indeed the case, the subgroup of $GL(6; \mathbb{R})$ that one is dealing with is isomorphic to $GL(3; \mathbb{C})$.

For a given direct sum decomposition of $\Lambda_2(\mathbb{R}^4) = \Lambda_{Re}(\mathbb{R}^4) \oplus \Lambda_{Im}(\mathbb{R}^4)$, any element $\alpha \in \Lambda_2(\mathbb{R}^4)$ can be expressed in the "complex" form:

$$\alpha = \alpha_R + *\alpha_I, \tag{1.5}$$

in which the terms in the sum represent the projections of $\alpha$ according to the direct sum, and $\alpha_I$ is the isomorphic element of $\Lambda_{Re}(\mathbb{R}^4)$ that corresponds to the imaginary projection.

One can now define the action of $\mathbb{C}$ on $\Lambda_2(\mathbb{R}^4)$ directly by analogy with its action on $\mathbb{C}^3$:

$$(\alpha + i\beta)(\alpha_R + *\alpha_I) = (\alpha\alpha_R - \beta\alpha_I) + *(\beta\alpha_I + \alpha\alpha_R). \tag{1.6}$$

If we exhibit a bivector $\alpha$ as a column vector $[\alpha_R, *\alpha_R]^T$ then complex scalar multiplication by $\alpha + i\beta$ can be also expressed as the action of a real invertible 2×2 matrix:

$$\begin{bmatrix} \alpha & \beta \\ -\beta & \alpha \end{bmatrix} = \begin{bmatrix} \alpha & 0 \\ 0 & \alpha \end{bmatrix} + \begin{bmatrix} 0 & \beta \\ -\beta & 0 \end{bmatrix} = \alpha I_2 + \beta J. \tag{1.7}$$

The set of all such matrices (except 0) given matrix multiplication defines a faithful real representation of the multiplicative group $(\mathbb{C}^*, \times)$ of non-zero complex numbers. The polar form of a complex number basically amount to the expression of such a matrix as a positive scalar constant equal to the modulus of $\alpha + i\beta$ – which is, of course, the square-root of the determinant of the matrix (1.7) – times a duality rotation matrix, which we now discuss.

Actually, as we shall discuss in more detail shortly, there is a way of distinguishing two distinct types of 3-planes in $\Lambda_2(\mathbb{R}^4)$ depending on whether all of the elements in the 3-plane do or do not admit the same common exterior factor. We shall use this distinction to determine the "realness" or "imaginariness" of a 3-plane in $\Lambda_2(\mathbb{R}^4)$.

*c. Duality rotations and duality planes.* There is another way of defining complex scalar multiplication on $\Lambda_2(\mathbb{R}^4)$ that basically amounts to considering the polar form of the complex scalar. Since we already have an action of $\mathbb{R}$ by scalar multiplication, if we express an element $\alpha \in \mathbb{C}$ in polar form as $\alpha = r(\cos\theta + i\sin\theta)$ then we see that all that we have added is an action of $U(1)$ on $\Lambda_2(\mathbb{R}^4)$. If we imitate its action on $\mathbb{C}^3$, namely:

$$(\cos\theta + i\sin\theta)(x^i + i\, y^i) = (x^i \cos\theta - y^i \sin\theta) + i(x^i \sin\theta + y^i \cos\theta), \tag{1.8}$$

by using the * isomorphism in place of $i$ then if we define $\alpha = \alpha_R + *\alpha_I$ we can define the action of $U(1)$ on bivectors by:

$$(\cos\theta + i\sin\theta)(\alpha_R + {}^*\alpha_I)$$
$$= (\cos\theta\,\alpha_R - \sin\theta\,\alpha_I) + {}^*(\sin\theta\,\alpha_I + \cos\theta\,\alpha_R), \tag{1.9}$$

which can also be represented as an action of $SO(2)$:

$$\begin{bmatrix} \cos\theta & -\sin\theta \\ \sin\theta & \cos\theta \end{bmatrix} \begin{bmatrix} \alpha_R \\ \alpha_I \end{bmatrix} = \begin{bmatrix} \cos\theta\,\alpha_R - \sin\theta\,\alpha_I \\ \sin\theta\,\alpha_I + \cos\theta\,\alpha_R \end{bmatrix}. \tag{1.10}$$

One refers to this action of $SO(2)$ on $\Lambda_2(\mathbb{R}^4)$ as the action of *duality rotations* on 2-forms. The pair of bivectors $\alpha$ and $^*\alpha$ span the same 2-plane as $\alpha_R$ and $^*\alpha_I$, and we refer to that plane as the *duality plane* of $\alpha$. If $\phi$ is a $\mathbb{C}$-linear functional on $\Lambda_2(\mathbb{R}^4)$ then a duality rotation of a 2-form $\alpha$ through an angle $\theta$ will produce a corresponding rotation of the value of $\phi(\alpha)$ in the complex plane.

Under a $\mathbb{C}$-linear isomorphism of $\Lambda_2(\mathbb{R}^4)$ with $\mathbb{C}^3$, one sees that the duality planes in $\Lambda_2(\mathbb{R}^4)$ correspond to the complex lines through the origin of $\mathbb{C}^3$. Hence, there is a one-to-one correspondence between the elements of the set of all duality planes in $\Lambda_2(\mathbb{R}^4)$ and the elements of $\mathbb{C}P^2$.

One sees immediately that the orbit of any bivector under all duality rotations spans its duality plane. As a result, one also sees that duality planes are invariant under all duality rotations, and, in particular, the * isomorphism itself. Hence, we see that any 3-plane $\Pi_3$ in $\Lambda_2(\mathbb{R}^4)$ that contains a duality plane $\Delta$ will have the property that $\Pi_3 \cap {}^*\Pi_3 = \Delta \neq 0$. Conversely, if a 3-plane $\Pi_3$ has the property that $\Pi_3 \cap {}^*\Pi_3 = \Delta \neq 0$ then $\Delta$ is a duality plane. This is because $\Delta$ could not be a line, since duality planes are orbits, and can only intersect completely or not at all. This implies:

Proposition:

*A 3-plane $\Pi_3$ in $\Lambda_2(\mathbb{R}^4)$ contains a duality plane $\Delta$ iff $\Pi_3 \cap {}^*\Pi_3 = \Delta \neq 0$.*

Note that it is impossible for any 3-plane $\Pi_3$ to satisfy $\Pi_3 \cap {}^*\Pi_3 = \Pi_3$; in fact, since any such intersection must contain only disjoint duality planes:

Proposition:

*If a $k$-plane $\Pi_k$ in $\Lambda_2(\mathbb{R}^4)$ has the property $\Pi_k \cap {}^*\Pi_k = \Pi_k$ (i.e., $^*\Pi_k = \Pi_k$) then $k$ is even.*

In general:

Proposition:

*For any $k$-plane $\Pi_k$ in $\Lambda_2(\mathbb{R}^4)$ the intersection $\Pi_k \cap {}^*\Pi_k$ is either zero or it consists of a direct sum of duality planes.*

We shall call a $k$-plane in $\Lambda_2(\mathbb{R}^4)$ *degenerate* if it contains any duality planes, and *non-degenerate* otherwise. Clearly, a $k$-plane $\Pi_k$ is (non-)degenerate iff $*\Pi_k$ is (non-)degenerate.

Since duality planes in $\Lambda_2(\mathbb{R}^4)$ are the $\mathbb{C}$-linear isomorphs of complex lines in $\mathbb{C}^3$ and one can express $\mathbb{C}^3$ as a direct sum of three linearly independent complex lines, one also has that $\Lambda_2(\mathbb{R}^4)$ can be expressed as a direct sum of three linearly independent duality planes. This is equivalent to saying that we can find a complex basis of bivectors for $\Lambda_2(\mathbb{R}^4)$ of the form $E_i$, $i = 1, 2, 3$ such that none of the duality planes intersect non-trivially. Such a complex basis would then define a real basis $E_i$, $*E_i$, $i = 1, 2, 3$ such that the 2-planes spanned by the pairs $\{E_i, *E_i\}$ would give a decomposition by duality planes.

*d. Complex bivectors and self-duality.* Perhaps because of the deep and compelling results that had been obtained in gauge field theory by the consideration of self-dual 2-forms on Riemannian manifolds, it was not surprising that relativity theory, which has long been trying to reconcile the geometric nature of gauge field theories, such as electromagnetism, with the geometric nature of gravitational field theory, pursued a parallel path of development regarding the role of self-duality in gravitational field theories (cf., [**4-6**]).

Of course, the immediate distinction to be made is that when one has defined a *-isomorphism of $\Lambda_2(\mathbb{R}^4)$ with itself that is of Riemannian type – i.e., $*^2 = +I$ – since the eigenvalues of the *-operator (viz., $\pm 1$) are then real, one can speak of a unique decomposition of $\Lambda_2(\mathbb{R}^4)$ into eigenspaces of *: $\Lambda_2^+(\mathbb{R}^4) \oplus \Lambda_2^-(\mathbb{R}^4)$, where $\Lambda_2^+(\mathbb{R}^4)$ is the space of *self-dual* bivectors, which have the positive eigenvalue, and $\Lambda_2^-(\mathbb{R}^4)$ is the space of *anti-self-dual* bivectors, which take the negative eigenvalue. In fact, the decomposition of any bivector $\alpha$ into a self-dual part $\alpha_+$ and an anti-self-dual part $\alpha_-$ is then effected by a simple polarization process that is completely analogous to (1.3).

As we pointed out in the last section, when one goes over to a *-operator of Lorentzian type the eigen-bivectors of * are not real since the eigenvalues are now $\pm i$. Although we decided above that the solution to that dilemma was to simply live with the non-uniqueness of the decomposition of $\Lambda_2(\mathbb{R}^4)$ into $\Lambda_{Re}(\mathbb{R}^4) \oplus \Lambda_{Im}(\mathbb{R}^4)$ and pursue the geometric consequences of such decompositions, we should point out that when uses complex bivectors, for which multiplication by $i$ does not produce a bivector that is foreign to the space, the eigenspace decomposition is well-defined and unique, as in the Riemannian case ([3]).

In order to account for the methodology of self-dual complex bivectors as it is commonly practiced, we need only establish the $\mathbb{C}$-linear isomorphism of $\mathbb{C}^3$ with the complex vector space of all such bivectors. This will then establish that the difference between using real bivectors and a given decomposition of $\Lambda_2(\mathbb{R}^4)$ into $\Lambda_{Re}(\mathbb{R}^4) \oplus \Lambda_{Im}(\mathbb{R}^4)$, versus using self-dual complex bivectors is simply a matter of choosing a representation for the slightly more abstract complex vector space $\mathbb{C}^3$.

---

[3] This is somewhat like the way that a rotation of the real plane has no non-zero eigenvector in two dimensions, but it does have a unique rotational axis in three-dimensions.

The *complexification* of $\Lambda_2(\mathbb{R}^4)$, which we denote by $\Lambda_2(\mathbb{R}^4) \otimes \mathbb{C}$, is a six-dimensional complex vector space that one obtains from $\Lambda_2(\mathbb{R}^4)$ by enlarging the space of scalars from $\mathbb{R}$ to $\mathbb{C}$. For instance, if $\mathbf{F} = \frac{1}{2} F^{\mu\nu} \mathbf{e}_\mu \wedge \mathbf{e}_\nu$ then we simply understand that the components $F^{\mu\nu}$ are complex numbers now. Hence, we could just as well write a complex $\mathbf{F}$ as:

$$\mathbf{F} = \tfrac{1}{2} F^{\mu\nu} \mathbf{e}_\mu \wedge \mathbf{e}_\nu + \tfrac{i}{2} G^{\mu\nu} \mathbf{e}_\mu \wedge \mathbf{e}_\nu, \tag{1.11}$$

in which the components are all real. Similarly, one could say that we have enlarged our real basis $\mathbf{e}_\mu \wedge \mathbf{e}_\nu$ by its imaginary counterparts $i\mathbf{e}_\mu \wedge \mathbf{e}_\nu$.

If one extends the *-isomorphism on $\Lambda_2(\mathbb{R}^4)$ to $\Lambda_2(\mathbb{R}^4) \otimes \mathbb{C}$ by demanding that it commute with $i$ (viz., $*\alpha = *\text{Re}(\alpha) + i\,*\text{Im}(\alpha)$) then one sees that the eigen-bivectors of * are the complex bivectors of the form:

$$\alpha = \mathbf{F} \pm i*\mathbf{F}. \tag{1.12}$$

Surprisingly, the *negative* sign refers to self-dual bivectors and the *positive* sign to the anti-self-dual ones. Hence, the polarization of elements of $\Lambda_2(\mathbb{R}^4) \otimes \mathbb{C}$ by Lorentzian duality that corresponds to the Riemannian case (1.8) is simply:

$$\alpha_+ = \tfrac{1}{2}(\alpha - i*\alpha), \qquad \alpha_- = \tfrac{1}{2}(\alpha + i*\alpha). \tag{1.13}$$

Of course, if we form these expressions without the factor of $i$ then we obtain real expressions that give us the decomposition of $\Lambda_2(\mathbb{R}^4)$ that is described by (1.3). It is clear that the decomposition of $\Lambda_2(\mathbb{R}^4) \otimes \mathbb{C}$ that is defined by (1.11) is simply the complexification of the decomposition of $\Lambda_2(\mathbb{R}^4)$ that is defined by (1.3); all that we have really done is to replace * with $i*$.

Now, the self-dual subspace of $\Lambda_2(\mathbb{R}^4) \otimes \mathbb{C}$ – as well as the anti-self-dual subspace, for that matter – is three-dimensional as a complex vector space; hence, it is $\mathbb{C}$-linear isomorphic to $\mathbb{C}^3$. One way of specifically defining that isomorphism is to first map the basis $\{\mathbf{e}_\mu \wedge \mathbf{e}_\nu, *(\mathbf{e}_\mu \wedge \mathbf{e}_\nu)\}$ for $\Lambda_2(\mathbb{R}^4)$ to the basis $\{E_i, i = 1, 2, 3\} = \{\tfrac{1}{2}(\mathbf{e}_\mu \wedge \mathbf{e}_\nu - i*(\mathbf{e}_\mu \wedge \mathbf{e}_\nu))$, all independent $\mu, \nu\}$ for the self-dual subspace of $\Lambda_2(\mathbb{R}^4) \otimes \mathbb{C}$; this establishes the $\mathbb{R}$-linear isomorphism. More specifically, this basis is:

$$E_i = \begin{cases} \tfrac{1}{2}(\mathbf{e}_0 \wedge \mathbf{e}_1 - i\mathbf{e}_2 \wedge \mathbf{e}_3), \\ \tfrac{1}{2}(\mathbf{e}_0 \wedge \mathbf{e}_2 - i\mathbf{e}_3 \wedge \mathbf{e}_1), \\ \tfrac{1}{2}(\mathbf{e}_1 \wedge \mathbf{e}_2 - i\mathbf{e}_0 \wedge \mathbf{e}_3). \end{cases} \tag{1.14}$$

We can see that the action of * on an element of $\Lambda_2(\mathbb{R}^4)$ corresponds to the multiplication of a self-dual complex bivector by $i$. Hence, the isomorphism takes * to multiplication by $i$, which makes the isomorphism $\mathbb{C}$-linear.

We observe, in passing, that another way of passing from * to $i*$ is to multiply the volume element $\mathcal{V}$ by $i$. When the volume element is consistent with a Lorentzian metric tensor $g$ – i.e., $\mathcal{V} = \sqrt{-\det(g)}\,\mathbf{e}_0 \wedge \mathbf{e}_1 \wedge \mathbf{e}_2 \wedge \mathbf{e}_3$ for an orthonormal frame $\mathbf{e}_\mu$ – this has the effect of replacing $\sqrt{-\det(g)}$ with $\sqrt{\det(g)}$, since $\det(g)$ is negative.

Since the difference between defining a complex structure on $\Lambda_2(\mathbb{R}^4)$ directly by way of * versus indirectly by way of an isomorphism with the self-dual complex bivectors is essentially a matter of convenience, although the representation seems to be convenient to the canonical formulation of gravitation and the study of gravitational radiation, nevertheless, in the discussion that follows − which is, of course, oriented towards electromagnetism − we shall take the direct route. Of course, insofar as gravitation is an unavoidable part of spacetime geometry that intersects the electromagnetic structure of spacetime in the structure of the light cones that common to both, eventually one must reconcile the two formalisms. Mostly, this seems to involve a lot of stray factors of $i$ that originate in the replacement $\mathcal{V} \mapsto i\mathcal{V}$.

*e. Scalar products on* $\Lambda_2(\mathbb{R}^4)$. When one works with complex vector spaces instead of real ones, one finds that a choice of unit volume element can serve to define geometric notions that are equivalent to a choice of scalar product on a real space. For instance, the canonical volume element $\varepsilon = \frac{1}{2} \varepsilon_{IJ} \theta^I \wedge \theta^J$ on $\mathbb{C}^2$ − i.e., the space of *2-spinors* − establishes a one-to-one correspondence between unit-volume 2-frames and complex 2×2 matrices of unit determinant, which then preserve the determinant of complex 2×2 matrices, i.e., elements of $\mathfrak{sl}(2;\mathbb{C})$, under conjugation, and there is a subspace of $\mathfrak{sl}(2;\mathbb{C})$ that one can identify with Minkowski space in such a way that the determinant of a matrix in that subspace equals the quadratic form that is associated with the corresponding vector in Minkowski space.

We shall now find that the vector space $\Lambda_2(\mathbb{R}^4)$, due to its special nature, can be given a natural real scalar product, and when one introduces a self-adjoint complex structure, it can be given a second real scalar product, as well as two complex ones, one of which is isometric to the complex Euclidian scalar product on $\mathbb{C}^3$ and the other of which is unitarily equivalent to the Hermitian scalar product on $\mathbb{C}^3$.

We give $\mathbb{R}^4$ a specific choice of unit volume element $\mathcal{V} \in \Lambda^4(\mathbb{R}^4)$, such as one might obtain from the canonical coframe. Such a 4-form can also be regarded as a linear functional on 4-vectors.

The natural scalar product on $\Lambda_2(\mathbb{R}^4)$ is defined by:
$$<\alpha, \beta> = \mathcal{V}(\alpha \wedge \beta). \qquad (1.15)$$

Since our * isomorphism was defined abstractly, and not by the usual Hodge duality, in this section, we must further assume that * is *self-adjoint*; i.e.:
$$\alpha \wedge {*\beta} = {*\alpha} \wedge \beta \qquad (1.16)$$

for all $\alpha, \beta \in \Lambda_2(\mathbb{R}^4)$. As a consequence, one has:
$$*\alpha \wedge *\beta = -\alpha \wedge \beta, \qquad (1.17)$$

as well. One can interpret this statement by saying that * takes $<\alpha, \beta>$ to $-<\alpha, \beta>$; i.e., it is an isometry of the scalar product $<.,.>$, up to sign.

Suppose $\Lambda_2(\mathbb{R}^4)$ has been given a specific choice of decomposition into a real subspace and an imaginary one. If $\alpha = \alpha_R + *\alpha_I$ then the scalar product of $\alpha$ with $\beta = \beta_R + *\beta_I$ takes the form:
$$<\alpha, \beta> = \mathcal{V}(\alpha_R \wedge \beta_R + *\alpha_I \wedge *\beta_I) + \mathcal{V}(\alpha_R \wedge *\beta_I + *\alpha_I \wedge \beta_R), \qquad (1.18)$$
namely:
$$<\alpha, \beta> = <\alpha_R, \beta_R> - <\alpha_I, \beta_I> + <\alpha_R, *\beta_I> + <*\alpha_I, \beta_R>. \qquad (1.19)$$

If $E_i$, $*E_i$, $i = 1, 2, 3$ is a complex symmetric frame for $\Lambda_2(\mathbb{R}^4)$ and $E^i$, $*E^i$, $i = 1, 2, 3$ is its reciprocal co-frame then (1.19) can be expressed by means of the doubly covariant tensor:

$$\mathcal{K} = \mathcal{K}_{ij} E^i \otimes E^j + \mathcal{K}_{i,j+3} E^i \otimes *E^j + \mathcal{K}_{i+3,j} *E^i \otimes E^j + \mathcal{K}_{i+3,j+3} *E^i \otimes *E^j, \quad (1.20)$$

with:

$$\mathcal{K}_{ij} = <E^i, E^j> = -\mathcal{K}_{i+3,j+3}, \qquad \mathcal{K}_{i,j+3} = <E^i, *E^j> = \mathcal{K}_{i+3,j}. \quad (1.21)$$

When this basis takes the "canonical" form $E_i = \mathbf{e}_0 \wedge \mathbf{e}_i$, $i = 1, 2, 3$, $*E_1 = \mathbf{e}_2 \wedge \mathbf{e}_3$, $*E_1 = \mathbf{e}_2 \wedge \mathbf{e}_3$, $*E_1 = \mathbf{e}_2 \wedge \mathbf{e}_3$, for $\mathcal{V} = \theta^0 \wedge \theta^1 \wedge \theta^2 \wedge \theta^3$, the component matrix of $\mathcal{K}$ becomes:

$$\mathcal{K} = \begin{bmatrix} 0 & \delta_{ij} \\ \hline \delta_{ij} & 0 \end{bmatrix}. \quad (1.21)$$

If we consider the complex-valued form:

$$<\alpha, \beta>_{\mathbb{C}} = [<\alpha_R, \beta_R> - <\alpha_I, \beta_I>] + i(<\alpha_R, \beta_I> + <\alpha_I, \beta_R>] \quad (1.22)$$

then we see that this corresponds to the complex Euclidian scalar product on $\mathbb{C}^3$ under a choice of isomorphism, a scalar product whose tensor form in a canonical basis is:

$$\mathcal{E} = \delta_{ij} E^i \otimes E^j - \delta_{ij} *E^i \otimes *E^j + i\delta_{ij}(E^i \otimes *E^j + *E^i \otimes E^j), \quad (1.23)$$

with the associated component matrix:

$$\mathcal{E} = \begin{bmatrix} \delta_{ij} & i\delta_{ij} \\ \hline i\delta_{ij} & -\delta_{ij} \end{bmatrix}. \quad (1.24)$$

A particularly important difference between the complex Euclidian scalar product and the real analogue is the existence of non-zero complex vectors of zero length – i.e., *isotropic* vectors. By isometry, there will be bivectors with this property, as well, and we shall discuss them later.

Although the scalar product (1.15) that we defined on $\Lambda_2(\mathbb{R}^4)$ depends only upon the choice of volume element on $\mathbb{R}^4$, not a choice of complex structure, one can also define another real scalar product using that latter structure, by way of:

$$(\alpha, \beta) = \mathcal{V}(\alpha \wedge *\beta) = <\alpha, *\beta>. \quad (1.25)$$

Since:

$$\alpha \wedge *\beta = -(\alpha_I \wedge \beta_R + \alpha_R \wedge \beta_I) + (\alpha_R \wedge *\beta_R - *\alpha_I \wedge \beta_I), \quad (1.26)$$

we see that for a basis $Ei$, $*Ei$, $i = 1, 2, 3$, the tensor form of this scalar product is:

$$G = G_{ij} E^i \otimes E^j + G_{i,j+3} E^i \otimes *E^j + G_{i+1,j} *E^i \otimes E^j + G_{i+3,j+3} *E^i \otimes *E^j, \quad (1.27)$$

and if the basis is canonical then the component matrix takes the form:

$$G = \begin{bmatrix} \delta_{ij} & 0 \\ \hline 0 & -\delta_{ij} \end{bmatrix} = *\mathcal{K}, \qquad * \equiv \begin{bmatrix} 0 & \delta_{ij} \\ \hline -\delta_{ij} & 0 \end{bmatrix}, \quad (1.28)$$

which makes:

$$\mathcal{E} = G + i\mathcal{K}, \quad (1.29)$$

as expected.

We note that since:
$$(\alpha, \beta) = \langle \alpha, *\beta \rangle \tag{1.30}$$

both of the scalar products defined so far contain the same basic information, up to sign. Hence, if we associate this scalar product with the complex-valued form:
$$(\alpha, \beta)_{\mathbb{C}} = -(\langle \alpha_R, \beta_I \rangle + \langle \alpha_I, \beta_R \rangle) + i(\langle \alpha_R, \beta_R \rangle - \langle \alpha_I, \beta_I \rangle) \tag{1.31}$$

then we see that the two complex-valued forms that we have defined are related by:
$$(\alpha, \beta)_{\mathbb{C}} = i \langle \alpha, \beta \rangle_{\mathbb{C}} . \tag{1.32}$$

Consequently, for the complex geometry of $\Lambda_2(\mathbb{R}^4)$ we shall only concern ourselves with $\langle \alpha, \beta \rangle_{\mathbb{C}}$, since it subsumes the scalar products $\langle \alpha, \beta \rangle$ and $(\alpha, \beta)$ into a single complex orthogonal structure that is isometric to the one that is defined by (1.32).

We immediately note that since $\alpha \wedge *\beta = *\alpha \wedge \beta = -*\alpha \wedge *(*\beta)$, the scalar product (.,.) also enjoys the property that * is an isometry for it, up to sign.

The quadratic forms associated with the scalar products are:
$$\langle \alpha, \alpha \rangle = \langle \alpha_R, \alpha_R \rangle - \langle \alpha_I, \alpha_I \rangle + \langle \alpha_R, *\alpha_I \rangle + \langle *\alpha_I, \alpha_R \rangle \tag{1.33a}$$
$$(\alpha, \alpha) = -\langle \alpha_I, \alpha_R \rangle - \langle \alpha_R, \alpha_I \rangle + \langle \alpha_R, *\alpha_R \rangle - \langle *\alpha_I, \alpha_I \rangle, \tag{1.33b}$$
$$\langle \alpha, \alpha \rangle_{\mathbb{C}} = \langle \alpha, \alpha \rangle - i (\alpha, \alpha) . \tag{1.33c}$$

For a canonical basis one gets non-zero contributions only from:
$$\langle E_i, *E_i \rangle = \langle *E_i, E_i \rangle = (E_i, E_i) = \delta_{ij} . \tag{1.34}$$

Since we have a complex structure, as long as we have chosen a decomposition of $\Lambda_2(\mathbb{R}^4)$ into real and imaginary subspaces, we can also define the complex conjugate of a bivector $\alpha \in \Lambda_2(\mathbb{R}^4)$ in a manner that is consistent with the conjugate of a vector in $\mathbb{C}^3$. Hence, if $\alpha = \alpha_R + *\alpha_I$ then:
$$\alpha^{\dagger} = \alpha_R - *\alpha_I . \tag{1.35}$$

One can then define an inner product by:
$$((\alpha, \beta)) = (\alpha, \beta^{\dagger}). \tag{1.36}$$

The complex-valued form of this inner product then corresponds to a Hermitian inner product on $\mathbb{C}^3$. It expands to:
$$((\alpha, \beta))_{\mathbb{C}} = [(\alpha_R, \beta_R) + (\alpha_I, \beta_I)] + i [(\alpha_R, \beta_I) - (\alpha_I, \beta_R)]. \tag{1.37}$$

Its tensor is:
$$\mathcal{H} = \mathcal{H}_{ij} E^i \otimes E^j + \mathcal{H}_{i+3, j+3} *E^i \otimes *E^j + i(\mathcal{H}_{i, j+3} E^i \otimes *E^j + \mathcal{H}_{i+3, j} *E^i \otimes E^j), \tag{1.38}$$

with:
$$\mathcal{H}_{ij} = (E^i, E^j) = \mathcal{H}_{i+3, j+3}, \qquad \mathcal{H}_{i, j+3} = (E^i, *E^j) = -\mathcal{H}_{i+3, j} . \tag{1.39}$$

For a canonical basis, the component matrix then takes the form:
$$\mathcal{H} = \begin{bmatrix} \delta_{ij} & 0 \\ \hline 0 & \delta_{ij} \end{bmatrix} . \tag{1.40}$$

The quadratic form that this Hermitian structure defines is real for any bivector $\alpha$, namely:

$$((\alpha, \alpha))_{\mathbb{C}} = (\alpha_R, \alpha_R) + (\alpha_I, \alpha_I) . \tag{1.41}$$

Clearly, this scalar product is also preserved up to sign by *.

Since we pointed out that the definition of complex conjugate for bivectors depends upon the choice of decomposition of $\Lambda_2(\mathbb{R}^4)$ into real and imaginary subspaces, and this decomposition is not unique, one necessarily must address the extent to which a different choice of decomposition will give an inequivalent complex conjugate and Hermitian structure. As we shall see, the linear isomorphisms of $\Lambda_2(\mathbb{R}^4)$ that take one such decomposition to another one are the ones that commute with the * isomorphism, which give a subgroup of $GL(\Lambda_2(\mathbb{R}^4))$ that is isomorphic to $GL(3;\mathbb{C})$; hence, these isomorphisms will also take complex conjugates to complex conjugates.

*f. The Klein quadric.* An example of a quadric hypersurface that has a fundamental role in the mathematics of pre-metric electromagnetism is the *Klein quadric*. One starts by defining the *Plücker embedding* of the Grassmanian manifold $G_{2,4}$ of linear 2-planes in $\mathbb{R}^4$ into the (real) projectivized linear space $P\Lambda_2(\mathbb{R}^4)$ of equivalence classes of bivectors on $\mathbb{R}^4$ under non-zero scalar multiplication. In order to define this embedding, one chooses any two vectors **a** and **b** that span a 2-plane $\Pi_2$ and maps them to the line [**a**^**b**] in $\Lambda_2(\mathbb{R}^4)$ that is generated by the bivector **a**^**b**. Since a different choice of **a** and **b** would map to a bivector that differs from **a**^**b** by only the determinant of the matrix that takes the first frame in $\Pi_2$ to the second one, one sees that the map $\Pi_2 \mapsto [\mathbf{a} \wedge \mathbf{b}]$ is well-defined, and, in fact, an embedding.

Next, one notes that a line $[\alpha] \in P\Lambda_2(\mathbb{R}^4)$ is in the image of the Plücker embedding iff $\alpha$ is a decomposable bivector; i.e., it is of the form **a**^**b**. This, in turn, is true iff $\alpha$ has rank two, which is equivalent to either the statement that $\alpha \wedge \alpha = 0$ or the statement that the space of all $\mathbf{x} \in \mathbb{R}^4$ such that $\mathbf{x} \wedge \alpha = 0$ is two-dimensional.

If one defines the quadratic form on $\Lambda_2(\mathbb{R}^4)$ that is given by eq. (1.29a) then one sees that the image of the Plücker embedding is the real quadric hypersurface that this quadratic form defines in $P\Lambda_2(\mathbb{R}^4)$; one refers to this hypersurface as the *Klein quadric*.

One notes that the hyperplanes in $\Lambda_2(\mathbb{R}^4)$ that are tangent to this hypersurface are the ones that annihilate the 1-forms on $\Lambda_2(\mathbb{R}^4)$ that are dual to the elements of $\Lambda_2(\mathbb{R}^4)$ under this scalar product; i.e., the dual to $\alpha \in \Lambda_2(\mathbb{R}^4)$ is the 1-form $\alpha$ on $\Lambda_2(\mathbb{R}^4)$ such that $\alpha(\beta) = \langle \alpha, \beta \rangle$. Hence, the tangent space to the Klein quadric at $\alpha \in \Lambda_2(\mathbb{R}^4)$ is the vector space of all $\beta \in \Lambda_2(\mathbb{R}^4)$ such that $\alpha(\beta) = 0$. Since this vector space will be the same for all non-zero scalar multiples $\lambda \alpha$, one sees that one may regard this as a tangent space to $P\Lambda_2(\mathbb{R}^4)$, as well; in addition, one can regard the tangent space as defined by a line in the (real) projectivized dual space to $\Lambda_2(\mathbb{R}^4)$, namely, $P\Lambda^2(\mathbb{R}^4)$.

*g. The representation of 3-planes* in $\Lambda_2(\mathbb{R}^4)$. Now that we have discussed the representation of 2-planes in $\mathbb{R}^4$ by decomposable bivectors, we return to the question of decomposing $\Lambda_2(\mathbb{R}^4)$ into real and imaginary subspaces. We shall start with the fact that any three-dimensional subspace of $\Lambda_2(\mathbb{R}^4)$ must be composed of decomposable bivectors, since the non-zero elements must all be of rank two. (If this were not the case then one could form a non-zero 4-vector from the product of some pair of them, but this contradicts the fact that the subspace is three-dimensional.) We then observe that since any two bivectors in a three-dimensional subspace of $\Lambda_2(\mathbb{R}^4)$ must have a common exterior factor, there are two types of three-dimensional subspaces of $\Lambda_2(\mathbb{R}^4)$: the ones in which all elements have the *same* common exterior factor, and the ones in which they do not. We can then identify the former 3-planes as the injections of some 3-plane $\Pi_3$ in $\mathbb{R}^4$ by exterior multiplication with a vector **t** not contained in $\Pi_3$, and the latter 3-planes as the space of bivectors over $\Pi_3$. Hence, the type of decomposition that we shall deal with takes the form:

$$\Lambda_2(\mathbb{R}^4) = ([\mathbf{t}] \wedge \Pi_3) \oplus \Lambda_2(\Pi_3) \tag{1.42}$$

for some 3-plane $\Pi_3$ in $\mathbb{R}^4$.

The rest of this part of section **3** is largely concerned with rigorously establishing this fact.

If $\Pi_3$ is a 3-plane in $\mathbb{R}^4$ and **t** is any vector in $\mathbb{R}^4$ that is not in $\Pi_3$ then we can define a linear injection of $\Pi_3$ into $\Lambda_2(\mathbb{R}^4)$ by the map:

$$e_{\mathbf{t}} : \Pi_3 \to \Lambda_2(\mathbb{R}^4), \quad \mathbf{v} \mapsto \mathbf{t} \wedge \mathbf{v}. \tag{1.43}$$

The injectivity follows from the fact that if $\mathbf{t} \wedge \mathbf{v} = 0$ then $\mathbf{v} = \lambda \mathbf{t}$ for some scalar $\lambda$, but since we assumed that **t** is not in $\Pi_3$ the only possibility is that $\lambda = 0$. Actually, since any non-zero scalar multiple of **t** will map $\Pi_3$ to the same subspace $\Lambda_2(\mathbb{R}^4)$, we should think of this map as defined by the pair $\{[\mathbf{t}], \Pi_3\}$ where $[\mathbf{t}]$ is a line through the origin of $\mathbb{R}^4$ that is not contained in $\Pi_3$.

An immediate consequence of the definition of $e_{\mathbf{t}}$ is that the bivector that defines the image of **v** has rank two; hence, the image of $e_{\mathbf{t}}$ will lie in the Klein quadric. Of course, $e_{\mathbf{t}}$ cannot be surjective onto the Klein quadric, which is a five-dimensional submanifold in $\Lambda_2(\mathbb{R}^4)$.

There is another way of injecting $\Pi_3$ into $\Lambda_2(\mathbb{R}^4)$: one simply composes $e_{\mathbf{t}}$ with the * isomorphism. We shall denote this composition by the symbol $*_s = * \circ e_{\mathbf{t}}$; one can also say that $*_s(\mathbf{v}) = *(\mathbf{t} \wedge \mathbf{v})$. It behaves much like the Hodge isomorphism for Euclidian $\mathbb{R}^3$ that allows one to associate bivectors with cross products of vectors, as a result of the fact that:

Proposition:

*Under an injection of a 3-plane $\Pi_3$ that takes the form $e_{\mathbf{t}}$, one has:*
$$* \tilde{\Pi}_3 = *_s \Pi_3 = \Lambda_2(\Pi_3). \tag{1.44}$$

Proof:

Any element of $* \tilde{\Pi}_3$ is of the form $*(\mathbf{t} \wedge \mathbf{a})$ for some $\mathbf{a} \in \Pi_3$. Choose a frame $\mathbf{e}_i$, $i = 1, 2, 3$ for $\Pi_3$ such that $\mathcal{V} = \mathbf{t} \wedge \mathbf{e}_1 \wedge \mathbf{e}_2 \wedge \mathbf{e}_3$. Since $(\mathbf{t} \wedge \mathbf{e}_i) \wedge *(\mathbf{t} \wedge \mathbf{e}_i) = \mathcal{V}$ for all $i$, we see that:
$$*(\mathbf{t} \wedge \mathbf{a}) = a^i *(\mathbf{t} \wedge \mathbf{e}_i) = a^1 \mathbf{e}_2 \wedge \mathbf{e}_3 + a^2 \mathbf{e}_3 \wedge \mathbf{e}_1 + a^3 \mathbf{e}_1 \wedge \mathbf{e}_2. \tag{1.45}$$

The right-hand expression is clearly an element of $\Lambda_2(\Pi_3)$. Since * is a linear isomorphism and $\Lambda_2(\Pi_3)$ has the same dimension as $\tilde{\Pi}_3$, its image is all of $\Lambda_2(\Pi_3)$. Q.E.D.

Given a 3-plane $\Pi_3$ in $\mathbb{R}^4$ and a line [**t**] that is not in $\Pi_3$, one has a direct sum decomposition $\mathbb{R}^4 = [\mathbf{t}] \oplus \Pi_3$. Furthermore, if we denote the image of $\Pi_3$ under $e_\mathbf{t}$ by $\tilde{\Pi}_3$ then one has a corresponding direct sum decomposition:

$$\Lambda_2(\mathbb{R}^4) = \tilde{\Pi}_3 \oplus {}^*\tilde{\Pi}_3 \qquad (1.46)$$

as long as the 3-plane $\tilde{\Pi}_3$ is non-degenerate. Hence, we need to examine the question of whether a 3-plane $\tilde{\Pi}_3$ in $\Lambda_2(\mathbb{R}^4)$ that is the image of an injection of the form $e_\mathbf{t}$ can contain duality planes.

This would imply the existence of a bivector $\alpha \in \tilde{\Pi}_3$ of the form $\alpha = \mathbf{t} \wedge \mathbf{a}$ such that ${}^*\alpha \in \tilde{\Pi}_3$. Hence, ${}^*(\mathbf{t} \wedge \mathbf{a}) = \mathbf{t} \wedge \mathbf{b}$ for some $\mathbf{b} \in \Pi_3$. Hence:

$$<\alpha, {}^*\alpha> = (\alpha, \alpha) = \mathcal{V}(\mathbf{t} \wedge \mathbf{a} \wedge \mathbf{t} \wedge \mathbf{b}) = 0. \qquad (1.47)$$

Such bivectors – called *isotropic* bivectors – exist, and we shall discuss them in detail later due to their fundamental nature.

We now pose the question: if we are given a 3-plane $\tilde{\Pi}_3$ in $\Lambda_2(\mathbb{R}^4)$, does there always exist a 3-plane $\Pi_3$ and a vector **t** in $\mathbb{R}^4$ such that $\tilde{\Pi}_3$ is the image of $\Pi_3$ under the injection $e_\mathbf{t}$? It is clear that the bivectors of the 3-plane $\tilde{\Pi}_3$ must all have the same common exterior factor **t**, up to a scalar multiple. Hence, the question becomes: are there 3-planes in $\Lambda_2(\mathbb{R}^4)$ whose bivectors do not all have the same common exterior factor?

The key to envisioning the situation is to see that since rank-two bivectors in $\Lambda_2(\mathbb{R}^4)$ correspond to 2-planes in $\mathbb{R}^4$, when they have a common exterior factor that factor will generate a line of intersection of the two 2-planes. To say that all bivectors of a 3-plane in $\Lambda_2(\mathbb{R}^4)$ have a common exterior factor is to say that all of the 2-planes that they define intersect in a common line.

Proposition:

*If $\Pi_3$ is a 3-plane in $\mathbb{R}^4$ then $\Lambda_2(\Pi_3)$ is a 3-plane in $\Lambda_2(\mathbb{R}^4)$ whose elements have pairwise common exterior factors, but no exterior factor that is common to all of them.*

Proof: Since $\Lambda_2(\Pi_3)$ is a 3-plane in $\Lambda_2(\mathbb{R}^4)$, its elements must all be decomposable bivectors; hence, each such element defines a 2-plane in $\Pi_3$. Since $\Pi_3$ is 3-dimensional any two 2-planes must intersect in a subspace of dimension at least one. However, one can find three bivectors in $\Lambda_2(\Pi_3)$ whose 2-planes in $\Pi_3$ do not all intersect in a common line; for instance, if $\mathbf{e}_i$, $i = 1, 2, 3$ is a basis for $\Pi_3$ then the 2-planes associated with $\mathbf{e}_1 \wedge \mathbf{e}_2$, $\mathbf{e}_1 \wedge \mathbf{e}_3$, and $\mathbf{e}_2 \wedge \mathbf{e}_3$ intersect pairwise, but not in a line that is common to all three. Q.E.D.

Hence, we have the corollary:

Proposition:

*If a 3-plane $\tilde{\Pi}_3$ in $\Lambda_2(\mathbb{R}^4)$ takes the form of $[\mathbf{t}] \wedge \Pi_3$ for some 3-plane $\Pi_3$ in $\mathbb{R}^4$ then $*\tilde{\Pi}_3$ is a 3-plane that is not the image of any 3-plane in $\mathbb{R}^4$ by some injection of the form $e_\mathbf{t}$.*

Of course, one can still inject 2-planes in $\mathbb{R}^4$ into 2-planes in $*\tilde{\Pi}_3$ by way of an injection of the form $e_\mathbf{t}$ if one chooses a $\mathbf{t} \in \Pi_3$. The fact that the image of $\Pi_3$ under $e_\mathbf{t}$ cannot be three-dimensional in this case is due to the fact that $\mathbf{t} \wedge [\mathbf{t}] = 0$.

We call any nondegenerate 3-plane in $\Lambda_2(\mathbb{R}^4)$ that is of the form $\mathbf{t} \wedge \Pi_3$ *real* and any 3-plane of the form $*(\mathbf{t} \wedge \Pi_3) = \Lambda_2(\Pi_3)$ *imaginary*. Hence, we restate the decomposition of $\Lambda_2(\mathbb{R}^4)$ into a real and an imaginary 3-plane as:

$$\Lambda_2(\mathbb{R}^4) = \tilde{\Pi}_3 \oplus \Lambda_2(\Pi_3). \tag{1.48}$$

An immediate question to be addressed now is the extent to which a given decomposition of $\Lambda_2(\mathbb{R}^4)$ of this form determines a decomposition of $\mathbb{R}^4$ into some $\Pi_3$ and a $[\mathbf{t}]$ uniquely.

First, suppose we have such a decomposition as in (1.48) that results from a given choice of $[\mathbf{t}] \oplus \Pi_3$. Now choose another line $[\mathbf{t}']$ that does not lie in $\Pi_3$. If $\mathbf{t}'$ is a generator for the line $[\mathbf{t}']$ then one can decompose it into a part collinear with $[\mathbf{t}]$ and another contained in $\Pi_3$; i.e., $\mathbf{t}' = \tau \mathbf{t} + \mathbf{t}_\Pi$. If $\mathbf{v} \in \Pi_3$ then we have:

$$\mathbf{t}' \wedge \mathbf{v} = \tau \mathbf{t} \wedge \mathbf{v} + \mathbf{t}_\Pi \wedge \mathbf{v} \in \tilde{\Pi}_3 \oplus *\tilde{\Pi}_3. \tag{1.49}$$

Hence, if $\mathbf{t}'$ takes $\Pi_3$ to the same subspace as $\mathbf{t}$, i.e., $\mathbf{t}' \wedge \mathbf{v} \in \tilde{\Pi}_3$, then we must have that $\mathbf{t}_\Pi \wedge \mathbf{v} = 0$ for all $\mathbf{v} \in \Pi_3$; hence, we must have $\mathbf{t}_\Pi = 0$, which implies that $\mathbf{t}'$ is collinear with $\mathbf{t}$.

On the other hand, one obtains a very different situation when one fixes $[\mathbf{t}]$ and looks at other 3-planes $\Pi'_3$. Since any element $\mathbf{a}'$ of $\Pi'_3$ can be decomposed into $\mathbf{a} + \tau \mathbf{t}$ with $\mathbf{a} \in \Pi_3$ we see that this time, since $\mathbf{t} \wedge \mathbf{a}' = \mathbf{t} \wedge \mathbf{a} \in \tilde{\Pi}_3$, apparently, $\mathbf{t}$ takes *any* 3-plane in $\mathbb{R}^4$ that does not contain $[\mathbf{t}]$ to the *same* 3-plane in $\Lambda_2(\mathbb{R}^4)$.

We summarize the last two results in the:

Proposition:

*i) For a given 3-plane $\Pi_3$ in $\mathbb{R}^4$ and a 3-plane $\tilde{\Pi}_3$ in $\Lambda_2(\mathbb{R}^4)$, if there is a line $[\mathbf{t}]$ in $\mathbb{R}^4$ such that $[\mathbf{t}] \wedge \Pi_3 = \tilde{\Pi}_3$ then that line is unique.*

*ii) A given line $[\mathbf{t}]$ in $\mathbb{R}^4$ will take any 3-plane $\Pi_3$ in $\mathbb{R}^4$ that does not contain $[\mathbf{t}]$ to the same 3-plane $[\mathbf{t}] \wedge \Pi_3$ in $\Lambda_2(\mathbb{R}^4)$ as any other line $[\mathbf{t}']$ that does not lie in $\Pi_3$.*

This last proposition has a distinctly projective geometric character to it since it suggests that rather than deal with the Grassmanian manifold $G_{3,4}$ of 3-planes in $\mathbb{R}^4$, we could just as well deal with $\mathbb{R}P^3$, which is (non-canonically) diffeomorphic to it. Similarly, one could use the $\mathbb{R}P^{3*}$, which *is* canonically diffeomorphic to $G_{3,4}$ since every 3-plane $\Pi_3$ can be associated with a 1-form $\theta$ that annihilates it and is unique up to a non-zero scalar. Hence, we can regard a decomposition of $\mathbb{R}^4$ as being defined by a pair ([**t**], [$\theta$])$\in \mathbb{R}P^3 \times \mathbb{R}P^{3*}$ such that [$\theta$][**t**] $\neq$ 0; i.e., $\theta(\mathbf{t}) \neq 0$ for any non-zero generators of [**t**] and [$\theta$].

We point out that in twistor theory and complex relativity [**7-9**] the terminology for the two types of 3-planes that we have identified is *α-triplanes* for the injections of 3-planes in $\mathbb{R}^4$, and *β-triplanes* for the space of bivectors defined on some 3-plane in $\mathbb{R}^4$.

*h. Induced scalar products on k-planes in* $\mathbb{R}^4$. If $\Pi_k$ is a *k*-dimensional subspace of $\Lambda_2(\mathbb{R}^4)$ then one can give $\Pi_k$ a scalar product by restriction of a scalar product on $\Lambda_2(\mathbb{R}^4)$, and one calls it the *induced* scalar product. We shall mostly be concerned with the cases of $k$ = 2, 3, 4, especially in the cases of 2-planes and 3-planes that are the image of some 2-plane or 3-plane in $\mathbb{R}^4$ by an injection of the form $e_\mathbf{t}$. In this way, we hope to endow $\mathbb{R}^4$ with at least some of the geometry that $\Lambda_2(\mathbb{R}^4)$ possesses as a result of its various scalar products. We shall then address the issue of extending to a Lorentzian scalar product on $\mathbb{R}^4$ separately.

Of particular interest later is the fact that although the restriction of <.,.> to a duality 2-plane is of indefinite signature type – viz., (+1, −1), – nevertheless, there are 2-planes whose induced scalar product is indefinite that are not duality planes. For instance, if $E_i$, *$E_i$, $i$ = 1, 2, 3 is a complex orthonormal basis for $\Lambda_2(\mathbb{R}^4)$ then the plane spanned by $E_1$ and *$E_2$ would have this property.

Given a choice of scalar product on $\Lambda_2(\mathbb{R}^4)$ and an injection $e_\mathbf{t}: \Pi_k \to \Lambda_2(\mathbb{R}^4)$, ($k$= 2, 3) one can immediately define two scalar products on $\Pi_k$ by pull-back:

$$<\mathbf{v}, \mathbf{w}> = <\mathbf{t} \wedge \mathbf{v}, \mathbf{t} \wedge \mathbf{w}> = 0 \tag{1.50a}$$
$$(\mathbf{v}, \mathbf{w}) = (\mathbf{t} \wedge \mathbf{v}, \mathbf{t} \wedge \mathbf{w}) = <\mathbf{t} \wedge \mathbf{v}, *(\mathbf{t} \wedge \mathbf{w})> . \tag{1.50b}$$

Although we could try to pull back the Hermitian structure ((., .)), since $\mathbf{t} \wedge \mathbf{v}$ will always be "real," its conjugate bivector will be itself and the Hermitian structure pulls back to (1.50b). Furthermore, although the scalar product <.,.> plays an important role in $\Lambda_2(\mathbb{R}^4)$, we shall clearly not have use for its pullback to $\Pi_k$.

Now suppose we have 3-plane $\Pi_3$ and a line [**t**] that is not contained in $\Pi_3$. The scalar product (.,.) on $\Lambda_2(\mathbb{R}^4)$, which is of signature type (3, 3), may or may not define a Euclidian structure on $\Pi_3$. Since the character of the injection is mostly due to one's choice of [**t**], it is illuminating to examine the possible induced scalar products on $\Pi_3$ as we vary [**t**].

Suppose $\mathbf{e}_\mu$, $\mu$ = 0, 1, 2, 3 is a basis for $\mathbb{R}^4$ such that $E_i = \mathbf{e}_0 \wedge \mathbf{e}_i$, *$E_i$, $i$ = 1, 2, 3 is a canonical complex basis for $\Lambda_2(\mathbb{R}^4)$, so $(E_i, E_j) = \delta_{ij} = - (*E_i, *E_j)$. As it turns out, we can obtain all possible types of induced scalar scalar products by letting $\mathbf{t} = \mathbf{e}_0 + \varepsilon\mathbf{e}_1$, and $\Pi_3$ = span{$\mathbf{e}_1, \mathbf{e}_2, \mathbf{e}_3$} and letting $e$ range from 0 to $+\infty$. One then gets the injected basis for [**t**] $\wedge \Pi_3$:

$$\tilde{E}_1 = \mathbf{t} \wedge \mathbf{e}_1 = E_1, \quad \tilde{E}_2 = \mathbf{t} \wedge \mathbf{e}_2 = E_2 + \varepsilon * E_3, \qquad \tilde{E}_3 = \mathbf{t} \wedge \mathbf{e}_3 = E_3 - \varepsilon * E_2, \qquad (1.51)$$

corresponding to which the component matrix of the tensor associated with the induced scalar product is:

$$g = (\tilde{E}_i, \tilde{E}_j) = \begin{bmatrix} 1 & 0 & 0 \\ 0 & 1-\varepsilon^2 & 0 \\ 0 & 0 & 1-\varepsilon^2 \end{bmatrix}. \qquad (1.52)$$

It is then clear that there are three basic possibilities depending upon the value of $\varepsilon$:

*i*) $0 \le \varepsilon < 1$: In this case, $g$ is basically positive definite. We then call the injection $e_\mathbf{t}$ *Euclidian* and the line [$\mathbf{t}$], *timelike*. Hence, a timelike line in $\mathbb{R}^4$ injects a 3-plane that does not contain it to a real 3-plane in $\Lambda_2(\mathbb{R}^4)$.
*ii*) $\varepsilon = 1$: In this case, $g$ is doubly degenerate. We then call the line [$\mathbf{t}$], *lightlike*.
*iii*) $\varepsilon > 1$: Now, $g$ becomes doubly negative definite. In such a case, we call [$\mathbf{t}$] *spacelike*.

Hence, we have succeeded in recovering the usual causal trichotomy of lines in Minkowski space from the geometry of $\Lambda_2(\mathbb{R}^4)$ without introducing a full Lorentzian structure on $\mathbb{R}^4$ yet. We shall also use the terms timelike, lightlike, and spacelike to describe any vector that generates a line of that character.

One notes that had we chosen a different set of frame members for $\mathbf{t}$ and $\Pi_3$, we would have obtained the same results up to permutation. For instance, if we use $\mathbf{t} = \mathbf{e}_1 + \varepsilon \mathbf{e}_0$, and $\Pi_3 = \text{span}\{\mathbf{e}_0, \mathbf{e}_2, \mathbf{e}_3\}$ then the new injected basis is $-\tilde{E}_1$, $*\tilde{E}_3 + \varepsilon \tilde{E}_2$, $-*\tilde{E}_2 + \varepsilon \tilde{E}_3$ and the corresponding component matrix is diag($1, \varepsilon^2 - 1, \varepsilon^2 - 1$), which presents the same set of possibilities, but for different values of $\varepsilon$.

Of particular interest is the case in which $\varepsilon = 1$, since one also notes that in this case one has $\tilde{E}_2 = *\tilde{E}_3$. Hence, the 2-plane that $\tilde{E}_2$ and $\tilde{E}_3$ span in $\Lambda_2(\mathbb{R}^4)$ is a duality plane. The fact that the restriction of the scalar product (.,.), as well as <.,.>, to such a duality plane vanishes in the lightlike case will become self-evident when we examine the case of isotropic bivectors. We can then say that [$\mathbf{t}$] is lightlike iff it takes some 2-plane in $\mathbb{R}^4$ to a duality plane in $\Lambda_2(\mathbb{R}^4)$.

First, we observe that if [$\mathbf{t}$] is timelike and our basis {$\mathbf{e}_1, \mathbf{e}_3, \mathbf{e}_3$} for $\Pi_3$ is orthonormal for induced scalar product then the 2-plane spanned by $\mathbf{e}_2$ and $\mathbf{e}_3$ is normal to the unit vector $\mathbf{e}_1$. Hence, the image of this 2-plane under the injection by the lightlike vector $\mathbf{l} = \mathbf{t} + \mathbf{e}_1$ is a duality plane in $\Lambda_2(\mathbb{R}^4)$. More generally, by rotating the vector $\mathbf{e}_1$ in $\Pi_3$, we see that:

Proposition:

*An injection of the form $e_\mathbf{l}$ for a lightlike $\mathbf{l}$ takes a tangent plane to the unit sphere in $\Pi_3$ to a duality plane in $\Lambda_2(\mathbb{R}^4)$.*

In the next part of this study, we shall consider the opposite process.

*i. Isotropic bivectors.* As we observed above, the decomposable bivectors collectively define a real quadric in $\Lambda_2(\mathbb{R}^4)$ relative to the scalar product $\langle.,.\rangle$, namely, the Klein quadric. Now, we shall see that when $\Lambda_2(\mathbb{R}^4)$ is regarded as a complex vector space that has been given the complex Euclidian structure:

$$\langle \alpha, \alpha \rangle_{\mathbb{C}} = \langle \alpha, \alpha \rangle + i\, (\alpha, \alpha) \qquad (1.53)$$

the complex quadric that it defines, i.e., the set of all bivectors $\alpha$ such that:

$$\langle \alpha, \alpha \rangle_{\mathbb{C}} = 0 \qquad (1.54)$$

has considerable significance in both geometry and physics. By definition, such a bivector is called *isotropic*, or sometimes, *null*; in physics, such a bivector can represent a wavelike solution to the Maxwell equations.

Clearly, an isotropic bivector $\alpha$ must then satisfy both $\langle \alpha, \alpha \rangle = 0$ and $(\alpha, \alpha) = 0$. Since the *-isomorphism preserves both scalar products, up to sign, it takes isotropic bivectors to isotropic bivectors. Hence, $\alpha$ is isotropic iff $*\alpha$ is isotropic. One can directly verify that not only does the real line generated by an isotropic $\alpha$ consist exclusively of isotropic vectors, but so are all of the duality-rotated bivectors obtained from $\alpha$. Hence, all of the bivectors in the duality plane through $\alpha$ are isotropic. One then calls the duality plane of $\alpha$ an *isotropic 2-plane* in $\Lambda_2(\mathbb{R}^4)$. This is the real analogue of the more concise statement in $\mathbb{C}^3$ that if a vector is isotropic with respect to $\langle.,.\rangle_{\mathbb{C}}$ then, by the homogeneity of that condition, so are all of its complex scalar multiples.

More to the point, since the group $SO(3; \mathbb{C})$, by definition, preserves the value $\langle \alpha, \alpha \rangle_{\mathbb{C}}$ for all of the transformations of $\alpha$ that it contains, one also finds that the orbit of any isotropic bivector under the action of $SO(3; \mathbb{C})$ is isotropic. The natural question to ask is then: "what is the isotropy subgroup of this orbit?"

We already know that the group $SL(2; \mathbb{R})$ fixes any decomposable bivector. $SL(2; \mathbb{R})$ is not a subgroup of $SO(3; \mathbb{C})$, but one of its subgroups is, namely, $SO(2; \mathbb{R}) \cong U(1)$. The resulting homogeneous space $SO(3; \mathbb{C})/U(1)$ is four-dimensional as a real manifold.

As a complex hypersurface in a space that is $\mathbb{C}$-linearly isomorphic to $\mathbb{C}^3$, the algebraic set $\mathcal{I}$ of isotropic elements is a two-dimensional quadric; hence, when considered as a real algebraic set, it is four-dimensional. We shall now show that it is diffeomorphic to $\mathbb{R} \times S^1 \times S^2$; hence, as a submanifold of the real projectivization of $\Lambda_2(\mathbb{R}^4)$ it is diffeomorphic to $S^1 \times S^2$.

One notes that when a decomposition $\Lambda_2(\mathbb{R}^4) = \tilde{\Pi}_3 \oplus *\tilde{\Pi}_3$ is canonical for the scalar product $\langle \alpha, \beta \rangle$ one can form elements that satisfy $\langle \alpha, \alpha \rangle = \langle \alpha_R, \alpha_R \rangle - \langle \alpha_I, \alpha_I \rangle = 0$ by means of all bivectors of the form $\alpha_R \pm *\alpha_R$, where $\alpha_R \in \tilde{\Pi}_3$. However, they do not satisfy $(\alpha, \alpha) = 0$ since $(\alpha, \alpha) = \pm 2\langle \alpha_R, \alpha_R \rangle$, in this case. Now, any orthogonal transformation $A$ of $\tilde{\Pi}_3$ takes $\alpha_R$ to another bivector $A\alpha_R$ that still has the same norm, which means that $\alpha_R \pm *A\alpha_R$ still satisfies $\langle \alpha, \alpha \rangle = 0$. Hence, if $A$ takes $\alpha_R$ to an element that is perpendicular to the bivector $\alpha_R \pm *A\alpha_R$ will also satisfy $(\alpha, \alpha) = \pm 2\langle \alpha_R, A\alpha_R \rangle = 0$. For any non-zero $\alpha_R \in \tilde{\Pi}_3$, the set of possible bivectors that will define an isotropic element of the form $\alpha_R \pm {}^*A\alpha_R$ is then parameterized by the circle of radius-squared $\langle\alpha_R, \alpha_R\rangle$ in the 2-plane in $\tilde{\Pi}_3$ that is perpendicular to $\alpha_R$. Since the space of all $\alpha_R$ is three-dimensional, this gives us our four-dimensional space $\mathscr{I}$ of isotropic bivectors, which is then seen to be diffeomorphic to $\mathbb{R} \times S^1 \times S^2$ and when one passes to the real projectivization of $\Lambda_2(\mathbb{R}^4)$, it has the topology of $S^1 \times S^2$.

The most geometrically intuitive way to examine the character of isotropic bivectors is to use the fact that the first condition $\langle\alpha, \alpha\rangle = 0$ says that any isotropic bivector is decomposable. Hence, $\alpha$ is associated with a 2-plane $\Pi_2$ in $\mathbb{R}^4$ and $\alpha = \mathbf{a} \wedge \mathbf{b}$ for some (non-unique) choice of vectors $\mathbf{a}, \mathbf{b} \in \Pi_2$. One must also have that $*\alpha$ is decomposable, so it, too, is associated with a 2-plane $\Phi_2$ in $\mathbb{R}^4$ and can be given the form $*\alpha = \mathbf{c} \wedge \mathbf{d}$ for some (non-unique) choice of vectors $\mathbf{c}, \mathbf{d} \in \Phi_2$. The second condition $(\alpha, \alpha) = \langle\alpha, *\alpha\rangle = 0$ can be taken to imply that the 2-planes $\Pi_2$ and $\Phi_2$ intersect non-trivially. Now, their intersection cannot be two-dimensional or else one would have $\alpha = *\alpha$, which is not possible. Hence, one concludes the crucial result:

Proposition:

*Any isotropic bivector in $\Lambda_2(\mathbb{R}^4)$ is associated with a unique lightlike line in $\mathbb{R}^4$.*

Since we have already defined lightlike lines to be ones that take tangent spaces to the unit sphere in a 3-plane that does not contain the line, we need to check that part of our assertion.

We can say more about the nature of an isotropic bivector if we choose a vector $\mathbf{l}$ in that line of intersection $\Lambda$ between the 2-planes $\Pi_2$ and $\Phi_2$, and note that one will then have a vector $\mathbf{a} \in \Pi_2$ and another one $\mathbf{b} \in \Phi_2$ (by abuse of notation) such that:
$$\alpha = \mathbf{l} \wedge \mathbf{a}, \qquad *\alpha = \mathbf{l} \wedge \mathbf{b}. \tag{1.55}$$

Clearly, the line of intersection associated with $\alpha$ and $*\alpha$ is generated by $\mathbf{l}$. Now let $\Pi_2$ be the plane that is spanned by $\mathbf{a}$ and $\mathbf{b}$, but does not contain $\mathbf{l}$. The injection defined by $\mathbf{l}$ takes any linear combination of $\mathbf{a}$ and $\mathbf{b}$ to a linear combination of $\alpha$ and $*\alpha$. Hence, this injection takes $\Pi_2$ to a duality plane, and we conclude that $\mathbf{l}$ is lightlike.

Now that we know that isotropic 2-planes exist in $\Lambda_2(\mathbb{R}^4)$, we naturally ask what the maximum dimension of an isotropic $k$-plane would be; that is, the maximum real dimension of a linear subspace of the quadric. It certainly cannot be four, since the quadric itself is not a linear subspace. Hence, we need only examine the possibility of isotropic 3-planes. However, such a 3-plane must decompose into a duality 2-plane and a line not contained in it, and since the duality plane through that line must also be in the 3-plane, the only way that this could happen is if the two duality planes intersected in an isotropic line. But the duality plane of that line would have to be contained in both duality planes, which means they would agree, which contradicts our original assumption. Hence, the maximum dimension for an isotropic $k$-plane in $\Lambda_2(\mathbb{R}^4)$ is two.

From the last proposition, there is clearly a well-defined map from $\mathscr{I}$ to $\mathbb{R}P^3$ that takes any isotropic bivector to its corresponding lightlike line. Although it cannot be a surjection, since not all lines are lightlike, one still wonders what the actual dimension of the image might be. Since each decomposable bivector is fixed by $SL(2; \mathbb{R})$ and when we expand to $GL(2; \mathbb{R})$, all such transformed bivectors will span the same 2-plane in $\mathbb{R}^4$, we see that the issue with isotropic bivectors is whether an element of $GL(2; \mathbb{R})$ that fixes the 2-plane of $\alpha$ will also fix the 2-plane of $*\alpha$. This will be true iff it is an element of the subgroup of $GL(2; \mathbb{R})$ that is isomorphic to $(\mathbb{C}^*, \times)$. One then has that the map from $\mathscr{I}$ to the lightlike lines in $\mathbb{R}P^3$ is essentially the orbit map $\mathscr{I} \to \mathscr{I}/\mathbb{C}^*$, which has a two-dimensional image, which is, in fact, diffeomorphic to $S^2$, since:

$$\mathscr{I}/\mathbb{C}^* = \mathbb{R} \times S^1 \times S^2 / \mathbb{R} \times S^1.$$

Now, the 2-sphere in $\mathbb{R}P^3$ is associated with a three-dimensional cone in $\mathbb{R}^4$ that consists of all lightlike lines through the origin. Hence, we have shown:

Proposition:

*The set of all lightlike lines in $\mathbb{R}^4$ that correspond to the set of all isotropic bivectors is a three-dimensional cone with the origin as vertex.*

We then have the right to call that cone the *light cone* that is defined by the complex Euclidian structure on $\Lambda_2(\mathbb{R}^4)$. Hence, that complex Euclidian structure has reproduced the conformal structure of Minkowski space, and we still have yet to introduce the Lorentzian metric into $\mathbb{R}^4$ explicitly.

Now, let us see how all of this relates to a choice of decomposition of $\Lambda_2(\mathbb{R}^4)$ into real and imaginary subspaces.

We already know that an isotropic bivector must take the form $\alpha = \mathbf{l} \wedge \mathbf{a}$, where $\mathbf{l}$ generates a lightlike line in $\mathbb{R}^4$. If we further assume that $\Pi_3$ is a 3-plane in $\mathbb{R}^4$ that contains $\mathbf{a}$ and $[\mathbf{t}]$ is a timelike line that is not contained in $\Pi_3$ then we can express $\mathbf{l}$ as $\mathbf{t} + \mathbf{n}$, where $\mathbf{n} \in \Pi_3$, and we have that a non-zero bivector $\alpha$ is isotropic iff it is expressible in the form:

$$\alpha = (\mathbf{t} + \mathbf{n}) \wedge \mathbf{a} = \mathbf{t} \wedge \mathbf{a} + \mathbf{n} \wedge \mathbf{a} \qquad (1.56)$$

for suitable vectors $\mathbf{a}, \mathbf{n} \in \Pi_3$ that make $\alpha \wedge *\alpha$ vanish.

Suppose we have an injection $e_\mathbf{t}$ of a 3-plane $\Pi_3$ into $\Lambda_2(\mathbb{R}^4)$ that is Euclidian and that $[\mathbf{l}]$ is a lightlike line. Any vector $\mathbf{l}$ that generates $[\mathbf{l}]$ can be decomposed into $\mathbf{t} + \mathbf{n}$, where $\mathbf{n} \in \Pi_3$. If we give $\Pi_3$ the induced scalar product $(\mathbf{a}, \mathbf{b}) = (\mathbf{t} \wedge \mathbf{a}, \mathbf{t} \wedge \mathbf{b})$ then we assume that $\mathbf{n}$ has unit norm. We can extend the scalar product to the rest of $\mathbb{R}^4$ by assuming that $\mathbf{t}$ is orthogonal to $\Pi_3$ and has a norm-squared of $-1$. Hence, if $\mathbf{a}, \mathbf{b} \in \mathbb{R}^4$ take the form $\mathbf{c} + \alpha \mathbf{t}$, $\mathbf{d} + \beta \mathbf{t}$, we have:

$$(\mathbf{a}, \mathbf{b}) = (\mathbf{c}, \mathbf{d}) + \alpha\beta(\mathbf{t}, \mathbf{t}) = (\mathbf{c}, \mathbf{d}) - \alpha\beta. \qquad (1.57)$$

In particular, any vector of the form $\lambda \mathbf{l}$ satisfies:

$$(\lambda \mathbf{l}, \lambda \mathbf{l}) = \lambda^2 (\mathbf{n}, \mathbf{n}) - \lambda^2 = 0. \qquad (1.58)$$

Hence, the light cone is the quadric defined by this Lorentzian scalar product.

The fact that this scalar product is really defined only up to a conformal factor is easily seen to be traceable to the fact that our choice of $\mathbf{t}$ in the line $[\mathbf{t}]$ was entirely arbitrary up to a non-zero scalar factor. This implicitly determines which vectors in $\Pi_3$ that we regarded as having unit norm since we really have $\langle \mathbf{n}, \mathbf{n} \rangle = \langle \mathbf{t} \wedge \mathbf{n}, \mathbf{t} \wedge \mathbf{n} \rangle$. Hence, had we chosen any other vector instead of $\mathbf{t}$, say $\lambda \mathbf{t}$, the equation for the light cone (1.58), by its homogeneity, would still remain valid, although the unit sphere in $\Pi_3$ would be rescaled by the factor $\lambda$.

*j. Timelike, lightlike, and spacelike k-planes.* We can characterize timelike, lightlike, and spacelike directions in $\mathbb{R}^4$ in another way by first characterizing 3-planes in $\mathbb{R}^4$ in that manner. We basically use the fact that we can define the lightlike directions in $\mathbb{R}^4$ in terms of the isotropic bivectors. In conventional Minkowski space, one has the *result* that a $k$-plane ($1 < k < 4$) is spacelike, lightlike, or timelike according to whether it contains zero, one, or more than one lightlike direction. Here, we use that as the *definition* of such $k$-planes. Then we can define a line in $\mathbb{R}^4$ to be spacelike iff it is contained in a spacelike $k$-plane for some $k$ and timelike iff any $k$-plane that contains it has at least two lightlike directions.

Of particular interest are the 2-planes in $\mathbb{R}^4$, since any 2-plane $\Pi_2$ corresponds to a line $[\mathbf{a} \wedge \mathbf{b}]$ through the origin of $\Lambda_2(\mathbb{R}^4)$ that lies in the Klein quadric. Indeed, if one restricts oneself to unit-volume 2-frames $\{\mathbf{a}, \mathbf{b}\}$ in $\Pi_2$ then the resulting bivector $\mathbf{a} \wedge \mathbf{b}$ is the same for all choices of $\mathbf{a}$ and $\mathbf{b}$. Furthermore, a 2-plane can contain exactly zero, one, or two lightlike directions, and in the last case, the lightlike directions can take the form $[\mathbf{t} \pm \mathbf{n}]$, where $\mathbf{t}$ is timelike and $\mathbf{n}$ is spacelike. In such a case, one can then define a *null* 2-frame by way of the null vectors $\mathbf{l}_\pm = \mathbf{t} \pm \mathbf{n}$, and one also has:

$$\mathbf{l}_+ \wedge \mathbf{l}_- = (\mathbf{t} + \mathbf{n}) \wedge (\mathbf{t} - \mathbf{n}) = -2\, \mathbf{t} \wedge \mathbf{n}, \qquad (1.59)$$

which is clearly timelike.

If a bivector $\mathbf{F}$ has rank four then it can be expressed in the form $\mathbf{F} = \mathbf{t} \wedge \mathbf{a} + \mathbf{b} \wedge \mathbf{c}$, such that $\mathbf{t} \wedge \mathbf{a} \neq 0$ is timelike (i.e., real) and $\mathbf{b} \wedge \mathbf{c} \neq 0$ is spacelike (i.e., imaginary). Hence, $\{\mathbf{t}, \mathbf{a}\}$ is a 2-frame that spans a timelike 2-plane in $\mathbb{R}^4$ and $\{\mathbf{b}, \mathbf{c}\}$ is a 2-frame that spans a spacelike one. The planes are, moreover, complementary, since they must collectively span $\mathbb{R}^4$. Again, any other 2-frames in these planes that differ from the chosen ones by a transformation in $SL(2; \mathbb{R})$ will produce the same pair of bivectors under exterior multiplication, and thus, the same $\mathbf{F}$. Hence, one can associate a bivector of rank four with a unique pair of complementary 2-planes.

If $\mathbf{F}$ is isotropic then it takes the form $\mathbf{l} \wedge \mathbf{a}$ where $\mathbf{l}$ is lightlike and $\mathbf{a}$ is spacelike, so the 2-plane spanned by the 2-frame $\{\mathbf{l}, \mathbf{a}\}$ is lightlike. If one expresses $\mathbf{F}$ in the form $(\mathbf{t} + \mathbf{n}) \wedge \mathbf{a} = \mathbf{t} \wedge \mathbf{a} + \mathbf{n} \wedge \mathbf{a}$ then one sees that $\mathbf{F}$ is also associated with the pair of 2-planes spanned by $\{\mathbf{t}, \mathbf{a}\}$ and $\{\mathbf{n}, \mathbf{a}\}$, which intersect in the line $[\mathbf{a}]$.

If $\mathbf{F}$ is of rank 2, but not isotropic, then it is of the form $\mathbf{a} \wedge \mathbf{b}$ and the 2-plane spanned by the 2-frame $\{\mathbf{a}, \mathbf{b}\}$ is either timelike or spacelike. If we want to associate $\mathbf{F}$ with a pair of 2-planes then we could say that in this case the 2-planes are both the same.

Ultimately, we can say:

Proposition:

*Any non-zero bivector $\mathbf{F} = \mathbf{a} \wedge \mathbf{b} + \mathbf{c} \wedge \mathbf{d}$ on $\mathbb{R}^4$ can be associated with a unique pair of 2-planes in $\mathbb{R}^4$ that are spanned by the 2-frames $\{\mathbf{a}, \mathbf{b}\}$ and $\{\mathbf{c}, \mathbf{d}\}$.*
*Their intersection will be:*

  *i*) *The origin if* **F** *has rank four*,
  *ii*) *A line if* **F** *is isotropic*,
  *iii*) *A 2-plane if* **F** *has rank two, but is not isotropic.*

Of course, if we multiply **F** by a non-zero real scalar then the resulting 2-planes will still be the same.

  *k. From light cones in* $\mathbb{R}^4$ *to Lorentzian structures.* Suppose $\mathbb{R}^4 = [\mathbf{t}] \oplus \Pi_3$ and the injection $e_\mathbf{t}: \Pi_3 \to \Lambda_2(\mathbb{R}^4)$ is Euclidian. Since we already pointed out that the only thing that prevents us from defining a Lorentzian injection of $\mathbb{R}^4$ into $\Lambda_2(\mathbb{R}^4)$ is a choice of bivector in $*\tilde{\Pi}_3 = \Lambda_2(\Pi_3)$, we see that now we are closer to such a definition, since any choice of unit vector $\mathbf{n} \in \Pi_3$ defines a choice of real bivector $\mathbf{t} \wedge \mathbf{n}$ and thus, an imaginary bivector $*(\mathbf{t} \wedge \mathbf{n})$. Hence, it is tempting to define our injection of $\mathbb{R}^4$ into $\Lambda_2(\mathbb{R}^4)$ by way of:

$$e(\mathbf{v}) = \begin{cases} e_\mathbf{t}(\mathbf{v}) & \text{if } \mathbf{v} \in \Pi_3, \\ *(\mathbf{v} \wedge \mathbf{n}) & \text{if } \mathbf{v} \in [\mathbf{t}]. \end{cases} \qquad (1.60)$$

We note that if $\mathbf{v} = v^0 \mathbf{t}$ then $*(\mathbf{v} \wedge \mathbf{n}) = v^0 *(\mathbf{t} \wedge \mathbf{n})$ and if we wish to regard $v^0$ the imaginary component of $e(\mathbf{v})$ then we must necessarily have that $*(\mathbf{t} \wedge \mathbf{n})$ has unit norm in $\Lambda_2(\Pi_3)$, which implies that $\mathbf{n}$ must have unit norm in $\Pi_3$. However, this simply means that $\mathbf{n}$ must lie on the unit sphere in $\Pi_3$. Hence, any such choice of $\mathbf{n}$ will produce the same $v^0$, and the Lorentzian structures on $\mathbb{R}^4$ that are obtained by demanding that $e$ be a Lorentzian isometry will be independent of any such choice of $\mathbf{n}$.

  Explicitly, if the injection $e: \mathbb{R}^4 \to \Lambda_2(\mathbb{R}^4)$ is Lorentzian – i.e., the image takes the form $\mathbf{t} \wedge \Pi_3 \oplus *([\mathbf{t}] \wedge \mathbf{n})$ and the restriction of (.,.) to that image is Lorentzian − then the Lorentzian structure on $\mathbb{R}^4$ is defined by pull-back:

$$(\mathbf{v}, \mathbf{w}) = (e(\mathbf{v}), e(\mathbf{w})) . \qquad (1.61)$$

We should observe that really what we are doing is using the complex vector space $\Lambda_2(\mathbb{R}^4)$ to represent an embedding of $\mathbb{R}^4$ into $\mathbb{C}^3$ that makes the "time" coordinate imaginary, so the Lorentzian structure on Minkowski is derivable from the complex orthogonal structure on $\mathbb{C}^3$. Of course, this "imaginary time" convention has been around for ages, except that the introduction of the imaginary *i* was always regarded as a matter of convenience, not a matter of geometric necessity.

  **2. Subgroups of** $GL(\Lambda_2(\mathbb{R}^4))$. Now that we have examined some of the structures that one might impose upon the vector space $\Lambda_2(\mathbb{R}^4)$, such as a complex structure, various decompositions, and various scalar products, both real and complex, we can discuss how they define subgroups of $GL(\Lambda_2(\mathbb{R}^4))$, which is non-canonically isomorphic to $GL(6; \mathbb{R})$, and how those subgroups act on $\Lambda_2(\mathbb{R}^4)$. Furthermore, the subgroup that is isomorphic to $SO(3; \mathbb{C})$ not only pertains to the complex Euclidian structure on $\Lambda_2(\mathbb{R}^4)$, but is isomorphic to the connected component of the Lorentz group, moreover, and thus defines a way of deriving the Lorentzian structure on $\mathbb{R}^4$ from the complex Euclidian structure on $\Lambda_2(\mathbb{R}^4)$.

*a. Action of GL*(3; $\mathbb{C}$) *on* $\Lambda_2(\mathbb{R}^4)$. We must now confront a subtlety in the $\mathbb{C}$-linear isomorphism of $\mathbb{C}^3$ with $\Lambda_2(\mathbb{R}^4)$: Although any linear transformation that acts on $\mathbb{C}^3$ clearly commutes with multiplication by *i*, it is *not* true that every linear transformation that acts on $\Lambda_2(\mathbb{R}^4)$ commutes with *.

Relative to a *symmetric complex frame* for $\Lambda_2(\mathbb{R}^4)$ – i.e., one of the form $\{E_i, *E_i\}$ – since * takes $E_i$ to $*E_i$ and $*E_i$ to $-E_i$, the matrix of * takes the form:

$$* = \begin{bmatrix} 0 & I_3 \\ -I_3 & 0 \end{bmatrix}. \tag{2.1}$$

If we express an element $A \in GL(\Lambda_2(\mathbb{R}^4))$ as a block matrix relative to the decomposition $\Lambda_2(\mathbb{R}^4) = \Lambda_{\text{Re}}(\mathbb{R}^4) \oplus \Lambda_{\text{Im}}(\mathbb{R}^4)$ that is defined by this symmetric complex frame $\{E_i, *E_i\}$:

$$A = \begin{bmatrix} B & C \\ \hline D & E \end{bmatrix} \tag{2.2}$$

then (2.1) implies that *A* must have the form:

$$A = \begin{bmatrix} B & C \\ \hline -C & B \end{bmatrix} = \begin{bmatrix} B & 0 \\ \hline 0 & B \end{bmatrix} + * \begin{bmatrix} C & 0 \\ \hline 0 & C \end{bmatrix}. \tag{2.3}$$

One naturally notes the similarity between such matrices and the ones that represent complex scalar multiplication, as in (1.7).

Since an element $A \in GL(3; \mathbb{C})$ can be expressed as $A = B + iC$, where *B* and *C* are real invertible 3×3 matrices, we examine whether the assignment:

$$B \mapsto \begin{bmatrix} B & 0 \\ \hline 0 & B \end{bmatrix}, \quad C \mapsto * \begin{bmatrix} C & 0 \\ \hline 0 & C \end{bmatrix} \tag{2.4}$$

actually defines an isomorphism of *GL*(3; $\mathbb{C}$) with the subgroup of elements of $GL(\Lambda_2(\mathbb{R}^4))$ that commute with *. Since the bijection is clear, one need only verify that that this assignment preserves matrix products, a verification that is also straightforward.

*b. Complex frames on* $\Lambda_2(\mathbb{R}^4)$. The various types of frames that one defines on $\Lambda_2(\mathbb{R}^4)$ can also be described by a choice of $\mathbb{C}$-linear isomorphism of $\Lambda_2(\mathbb{R}^4)$ with $\mathbb{C}^3$.

Of course, any choice of (real) linear frame $\{E_i, F_i\}$ for $\Lambda_2(\mathbb{R}^4)$ will define such an isomorphism by way of $\{E_i, F_i\} \mapsto E_i + iF_i$. However, one will still needs to distinguish different orbits of the real linear frames in $\Lambda_2(\mathbb{R}^4)$ according to the action of *GL*(3; $\mathbb{C}$) on $\Lambda_2(\mathbb{R}^4)$. Indeed, one must consider the homogeneous space *GL*(6; $\mathbb{R}$)/*GL*(3; $\mathbb{C}$), whose cosets essentially describe inequivalent complex structures, if one regards a complex structure as a reduction of *GL*(6; $\mathbb{R}$) to *GL*(3; $\mathbb{C}$) and equivalence as defined by conjugation.

Since the matrices of the real representation of $GL(3; \mathbb{C})$ are the matrices of $GL(6; \mathbb{R})$ that commute with *, the complement clearly represents the matrices of $GL(6; \mathbb{R})$ that do not. The difference between the real dimension of $GL(6; \mathbb{R})$ and that of $GL(3; \mathbb{C})$ is eighteen, which is then the dimension of the homogeneous space $GL(6; \mathbb{R})/GL(3; \mathbb{C})$.

The map * then defines a particular class of 6-frames for $\Lambda_2(\mathbb{R}^4)$: those 6-frames that are the image of complex 3-frames in $\mathbb{C}^3$ of the form $\tilde{E}_i = E_i + iF_i$ under a given choice of isomorphism, namely 6-frames of the form $\{E_i, *F_i\}$, which we will call *complex* frames, and the particular case of $\{E_i, *E_i\}$ will be called *symmetric complex* frames, as we did already.

*c. SO(3; $\mathbb{C}$) and its isomorphism with $SO_0(3, 1)$.* When $\Lambda_2(\mathbb{R}^4)$ is given the complex Euclidian scalar product (1.22), one can reduce the action of $GL(3; \mathbb{C})$ to an action of $O(3; \mathbb{C})$, and with the introduction of a unit-volume element on $\Lambda_2(\mathbb{R}^4)$, to an action of $SO(3; \mathbb{C})$. Given an oriented complex orthonormal 3-frame $\{E_i, *F_i\}$ in $\Lambda_2(\mathbb{R}^4)$, the elements of $SO(3; \mathbb{C})$ can then be put into one-to-one correspondence with the other oriented complex orthonormal 3-frames in $\Lambda_2(\mathbb{R}^4)$.

There is an isomorphism of $SO(3; \mathbb{C})$ with $SO_0(3, 1)$, which is the identity component of $SO(3, 1)$ and consists of oriented time-oriented Lorentz transformations of Minkowski space. This isomorphism is most transparent when one looks at the two Lie algebras of infinitesimal transformations.

The Lie algebra $\mathfrak{so}(3; \mathbb{C})$ consists of complex antisymmetric 3×3 matrices, which can then be represented by pairs $(\omega, \sigma)$ of real antisymmetric 3×3 matrices as $\omega + i\sigma$. Since this means that $\omega, \sigma \in \mathfrak{so}(3; \mathbb{R})$, if $J_i$, $i = 1, 2, 3$ are the standard basis matrices for $\mathfrak{so}(3; \mathbb{R})$ then we immediately see that the commutation rules for $\mathfrak{so}(3; \mathbb{C})$ are derived from those of $\mathfrak{so}(3; \mathbb{R})$ by way of:

$$[J_i, J_j] = \varepsilon_{ijk} J_k, \qquad [J_i, iJ_j] = i\varepsilon_{ijk} J_k, \qquad [iJ_i, iJ_j] = -\varepsilon_{ijk} J_k. \tag{2.5}$$

If we express the Lorentz Lie algebra $\mathfrak{so}(3,1)$ as a vector space direct sum $\mathfrak{so}(3) \oplus \mathfrak{b}$ and give it the standard basis $J_i$, $K_i$, $i = 1, 2, 3$, where the $K_i$ span the subspace $\mathfrak{b}$ of infinitesimal boosts then we see that its commutation rules are:

$$[J_i, J_j] = \varepsilon_{ijk} J_k, \qquad [J_i, K_j] = \varepsilon_{ijk} K_k, \qquad [K_i, K_j] = -\varepsilon_{ijk} J_k. \tag{2.6}$$

Hence, the $\mathbb{R}$-isomorphism of the Lie algebras simply amounts to replacing the purely imaginary matrices $iJ_i$ with the infinitesimal boosts $K_i$, respectively. Of course, this is also related to the basic equations of hyperbolic geometry that give us that:

$$\sinh x = -i \sin ix, \qquad \cosh x = \cos ix. \tag{2.7}$$

To exhibit the isomorphism of $SO(3; \mathbb{C})$ with $SO_0(3, 1)$, we start by observing that when one has chosen an oriented orthogonal almost-complex 3-frame $\{E_i, *F_i\}$ for $\Lambda_2(\mathbb{R}^4)$ an element $C \in SO(3; \mathbb{C})$, being an element of $GL(3; \mathbb{C})$, as well, can be expressed in the block matrix form:

$$C = \begin{bmatrix} A & B \\ -B & A \end{bmatrix} = \begin{bmatrix} A & 0 \\ 0 & A \end{bmatrix} + * \begin{bmatrix} B & 0 \\ 0 & B \end{bmatrix}. \tag{2.8}$$

If we assume that the chosen frame is canonical and call the matrix of the complex scalar product $\mathscr{E}$ then $\mathscr{E}$ takes the form:

$$\mathscr{E} = \left[\begin{array}{c|c} I & iI \\ \hline iI & -I \end{array}\right], \qquad (2.9)$$

and the condition that $C$ be complex orthogonal – namely, $C^T\mathscr{E}C = \mathscr{E}$ – implies the following conditions on $A$ and $B$:

$$A^TA - B^TB = I, \qquad A^TB + A^TB = 0. \qquad (2.10)$$

As a consequence of these, we note that when $C$ is "real," i.e., $B = 0$, then we must have that $A \in SO(3; \mathbb{R})$. When $C$ is "pure imaginary" we must have that $B^TB = -I$, which can be interpreted as either the constraint:

$$B^{-1} = -B^T, \qquad (2.11)$$

or the statement that $iB$ is a 3×3 rotation matrix, which is essentially the statement that $B$ represents a Lorentz boost. This is the key to making the isomorphism between $SO_0(3, 1)$ and $SO(3; \mathbb{C})$ explicit: we must regard a boost *in a given direction* as a rotation *around that direction*, but through an *imaginary* angle.

Hence, if we note that any element of $SO_0(3, 1)$ can be expressed as a product $\tilde{A}\tilde{B}$, where $A \in SO(3; \mathbb{R})$ and $B$ is a boost then if we have chosen our frame $\mathbf{e}_\mu$, $\mu = 0, 1, 2, 3$ in $\mathbb{R}^4$ such that $\mathbf{e}_0$ generates $[\mathbf{t}]$ and $\mathbf{e}_i$ frames $\Pi_3$ then we can express $\tilde{A}$ in the form:

$$\tilde{A} = \begin{bmatrix} 1 & 0 \\ 0 & A \end{bmatrix}, \qquad (2.12)$$

in which $A$ is a 3×3 real orthogonal matrix. One then maps the matrix $A$ to the 6×6 real matrix:

$$\begin{bmatrix} A & 0 \\ 0 & A \end{bmatrix}. \qquad (2.13)$$

It is only slightly more involved to represent the boost as 3×3 real invertible matrices. We simply use the aforementioned key to exhibit the elementary boosts as:

$$B_x = \begin{bmatrix} 1 & 0 & 0 \\ 0 & \cosh\alpha & \sinh\alpha \\ 0 & \sinh\alpha & \cosh\alpha \end{bmatrix}, B_y = \begin{bmatrix} \cosh\alpha & 0 & \sinh\alpha \\ 0 & 1 & 0 \\ \sinh\alpha & 0 & \cosh\alpha \end{bmatrix},$$

$$B_z = \begin{bmatrix} \cosh\alpha & \sinh\alpha & 0 \\ \sinh\alpha & \cosh\alpha & 0 \\ 0 & 0 & 1 \end{bmatrix}, \qquad (2.14)$$

and then embed them in 6×6 real matrices as:

$$*\begin{bmatrix} B & 0 \\ 0 & B \end{bmatrix} = \begin{bmatrix} 0 & B \\ -B & 0 \end{bmatrix}. \qquad (2.15)$$

We note that the product of two such boosts is then:

$$\begin{bmatrix} 0 & B_1 \\ -B_1 & 0 \end{bmatrix} \begin{bmatrix} 0 & B_2 \\ -B_2 & 0 \end{bmatrix} = \begin{bmatrix} -B_1 B_2 & 0 \\ 0 & -B_1 B_2 \end{bmatrix}, \tag{2.16}$$

and, from (2.14), we see that $-B_1 B_2$ is a rotation since $(-B_1 B_2)(-B_1 B_2)^T = I$. The product of two rotations is clearly another rotation, and when we multiply a rotation and a boost, we get:

$$\begin{bmatrix} A & 0 \\ 0 & A \end{bmatrix} \begin{bmatrix} 0 & B \\ -B & 0 \end{bmatrix} = \begin{bmatrix} 0 & AB \\ AB & 0 \end{bmatrix}, \tag{2.17}$$

which is seen to be a boost since $AB(AB)^T = -I$.

Hence, we have established that there is a one-to-one correspondence between 4×4 real proper orthochronous Lorentz matrices, when factored into the product $AB$ of a rotation and a boost, and real 6×6 matrices of the form (2.8), a correspondence that preserves the product of matrices, moreover. This gives us the explicit isomorphism of $SO_0(3, 1)$ with $SO(3; \mathbb{C})$ by means of the decomposition $\mathbb{R}^4 = [\mathbf{t}] \oplus \Pi_3$ and the corresponding decomposition $\Lambda_2(\mathbb{R}^4) = \tilde{\Pi}_3 \oplus {}^*\tilde{\Pi}_3$.

*d. Definition of a Lorentzian structure on* $\mathbb{R}^4$. This isomorphism can also be interpreted as a one-to-one correspondence between oriented complex-orthogonal 3-frames in $\mathbb{C}^3$ with oriented, time-oriented Lorentzian frames in $\mathbb{R}^4$. For instance, if one frames $\mathbb{R}^4$ with $\mathbf{e}_\mu$, $\mu = 0, 1, 2, 3$, and $\Lambda_2(\mathbb{R}^4)$ with $\{\mathbf{e}_0 \wedge \mathbf{e}_i, {}^*(\mathbf{e}_0 \wedge \mathbf{e}_i), i = 1, 2, 3\}$ then the frame $[AB]^\mu_\nu \mathbf{e}_\mu$ in $\mathbb{R}^4$ goes to the frame:

$$\begin{bmatrix} \mathbf{e}_0 \wedge \mathbf{e}_i & | & {}^*(\mathbf{e}_0 \wedge \mathbf{e}_i) \end{bmatrix} \begin{bmatrix} \tilde{A}^j_i & \tilde{B}^j_i \\ -\tilde{B}^j_i & \tilde{A}^j_i \end{bmatrix} =$$
$$\begin{bmatrix} \tilde{A}^j_i (\mathbf{e}_0 \wedge \mathbf{e}_i) - {}^*(\mathbf{e}_0 \wedge \tilde{B}^j_i \mathbf{e}_i) & | & \tilde{B}^j_i (\mathbf{e}_0 \wedge \mathbf{e}_i) + {}^*(\mathbf{e}_0 \wedge \tilde{A}^j_i \mathbf{e}_i) \end{bmatrix}. \tag{2.18}$$

Of course, it is probably simpler to think of this frame as being defined by the components of the 6×6 real matrix that produces it from the initial frame.

It is relatively straightforward to reverse the process if one wishes to start with a complex orthogonal 6-frame $\{E_i, {}^*E_i\}$ for $\Lambda_2(\mathbb{R}^4)$, since one of the two 3-frames, $E_i$ or ${}^*E_i$, will have a common factor $\mathbf{t}$ to all three 2-vectors, $\mathbf{t} \wedge \mathbf{e}_1, \mathbf{t} \wedge \mathbf{e}_2, \mathbf{t} \wedge \mathbf{e}_3$. From $\mathbf{t}$, we get $\mathbf{e}_0$, and from $\mathbf{e}_i$, we get the spacelike members of the frame. From the complex orthogonality of the 6-frame and the isomorphism above, we infer the Lorentz-orthogonality of the resulting 4-frame $\{\mathbf{e}_0, \mathbf{e}_i\}$ in $\mathbb{R}^4$.

The significance of the result that we just derived is in the fact that we did not actually introduce a Lorentzian structure on $\mathbb{R}^4$ at any point in our argument, so the reduction of $GL(4; \mathbb{R})$ to $SO_0(3, 1)$ was a by-product of the complex orthogonal structure on $\Lambda_2(\mathbb{R}^4)$, which was derived from the volume element on $T(\mathbb{R}^4)$ and the complex structure on $\Lambda_2(\mathbb{R}^4)$ ([4]).

---

[4] Perhaps we have arrived at what John Archibald Wheeler might call "metric without metric."

Once one has identified a class of Lorentzian orthonormal frames on $\mathbb{R}^4$ it is straightforward to define a Lorentzian scalar product $\eta$, since it will necessarily take the form:

$$\eta = \eta_{\mu\nu} \theta^\mu \otimes \theta^\nu, \qquad \eta_{\mu\nu} = \text{diag}\,(1, -1, -1, -1) \qquad (2.19)$$

whenever $\theta^\mu$ is the coframe that is reciprocal to a Lorentzian frame $\mathbf{e}_\mu$.

Of course, since the Lorentzian orthonormality of $\mathbf{e}_\mu$, $\mu = 0, 1, 2, 3$ is defined by the complex orthonormality of its image in $\Lambda_2(\mathbb{R}^4)$, namely, $E_i = \mathbf{e}_0 \wedge \mathbf{e}_i$, $*E_i$, $i = 1, 2, 3$, relative to the scalar product whose matrix is $\mathscr{E}$, it would be more illuminating to see how one could relate the components of $\eta$ in an arbitrary frame to the components of $\mathscr{E}$.

This is actually quite straightforward, especially when one takes advantage of the $\mathbb{C}$-linear isomorphism of $\Lambda_2(\mathbb{R}^4)$ with $\mathbb{C}^3$, the action of $GL(3; \mathbb{C})$ on $\Lambda_2(\mathbb{R}^4)$, and the isomorphism of $SO(3; \mathbb{C})$ with $SO_0(3, 1)$. Hence, if $\{E_i, *E_i\}$ is an arbitrary – i.e., not necessarily orthonormal – frame on $\Lambda_2(\mathbb{R}^4)$ then we associate with the canonical frame $\varepsilon_i$, $i = 1, 2, 3$ on $\mathbb{C}^3$. We express the matrix of $\mathscr{E}$ relative to the frame $\{E_i, *E_i\}$ as $\mathscr{E}_{IJ}$, I, J = 1, …, 6, so it becomes the complex matrix $\tilde{\mathscr{E}}_{ij} = \mathscr{E}_{ij} + i\, \mathscr{E}_{i+3,j}$, $i = 1, 2, 3$ relative to $\varepsilon_i$. We make a non-orthogonal transformation $\tilde{\mathscr{E}}_{ij} \mapsto C^T \tilde{\mathscr{E}}_{ij} C = \delta_{ij}$ to the canonical frame for $\tilde{\mathscr{E}}_{ij}$ in $\mathbb{C}^3$, where $C \in GL(3; \mathbb{C})$. $C$ also acts on $\{E_i, *E_i\}$, and the effect is to transform $\mathscr{E}_{IJ}$ to its canonical form $\text{diag}(+1, +1, +1, -1, -1, -1)$, as well. This canonical frame in $\Lambda_2(\mathbb{R}^4)$ for $\mathscr{E}$ then corresponds to a canonical frame in $\mathbb{R}^4$ for our Lorentzian metric $g$; i.e., one for which $g_{\mu\nu} = \eta_{\mu\nu}$. To get the components of $g$ in the original, non-canonical, frame one then needs to invert the original transformation $C$ to the canonical frame.

**3. Pre-metric electromagnetism.** The basic goal of pre-metric electromagnetism is to formulate the laws that govern the electromagnetic field and its interaction with charged matter in a manner that does not introduce the spacetime metric tensor field *a priori*, but deduces its existence as a consequence of more general and fundamental properties of spacetime that pertain to setting up the machinery for the propagation of electromagnetic waves. If one follows the genesis and development of relativity theory historically, one notices that the theory really started in the theory of electromagnetism, abstracted the Minkowski scalar product $\eta$ from the form of the characteristic equation for linear wave propagation, and then generalized to the Lorentzian metric $g$ on spacetime whose curvature implied the existence of gravitation. At an elementary level, one could observe that when one does not dismiss the appearance of $c_0$ in the components of the spacetime Lorentzian metric tensor as an irrelevant constant, one sees that conceptually the theory of gravitation still contains traces of its roots in the theory of electromagnetism. Indeed, one would not otherwise assume that the propagation speed of a gravitational wave need necessarily be the same as the corresponding value for an electromagnetic wave, except that one already assumes that the laws of causality that are defined by the propagation of electromagnetic waves are the definitive ones that govern all other forms of wave motion.

*a. Pre-metric form of Maxwell's equations* [**12, 13**]. One notices that if one states the Maxwell equations for electromagnetic fields in the usual form:

$$dF = 0, \qquad \delta F = -*d*F = \frac{4\pi}{c} J \qquad (4.1)$$

then the only place in which the introduction of a metric is implicit is in the use of the Hodge *-isomorphism. One can further analyze this situation by pointing out that the Hodge *-isomorphism is really the composition of two isomorphisms: an isomorphism between 2-forms and 2-vector fields:

$$\iota \wedge \iota : \Lambda^2(M) \to \Lambda_2(M), \qquad \alpha \wedge b \mapsto \iota(\alpha) \wedge \iota(\beta) \qquad (4.2)$$

that is a by-product of the isomorphism of $\iota: T^*(M) \to T(M)$ that the metric defines, and the Poincaré duality isomorphism:

$$\#: \Lambda_2(M) \to \Lambda^2(M), \quad \mathbf{a} \wedge \mathbf{b} \mapsto i_{\mathbf{a} \wedge \mathbf{b}} \mathcal{V}, \qquad (4.3)$$

Furthermore, from the physical perspective we can say that the isomorphism (4.2), when combined with the factor $c/4\pi$, which we rewrite as $1/(4\pi\sqrt{\varepsilon_0 \mu_0})$ actually takes the form of an elementary *linear electromagnetic constitutive law* for the spacetime vacuum:

$$\chi: \Lambda^2(M) \to \Lambda_2(M), \qquad F \mapsto \tilde{\mathfrak{H}} . \qquad (4.4)$$

If we treat the field strength 2-form $F$ as distinct from the induction 2-vector field $\tilde{\mathfrak{H}}$, except that they are related by the constitutive law $\tilde{\mathfrak{H}} = \chi(F)$, and regard the electric four-current that serves as the source of the field as a vector field **J** then we can restate the Maxwell equations as:

$$dF = 0, \qquad \delta\tilde{\mathfrak{H}} \equiv \#^{-1}d\#\tilde{\mathfrak{H}} = 4\pi\mathbf{J}, \qquad \tilde{\mathfrak{H}} = \chi(F), \qquad (4.5)$$

in which the factor $c$ has been absorbed into the constitutive law.

*b. Linear electromagnetic constitutive laws* [**14**]. A choice of linear electromagnetic constitutive law on an orientable, oriented manifold can, under certain physical conditions, define an almost-complex structure on $\Lambda^2(M)$, at least indirectly. To get some feel for what to expect, let us first review some of the aspects of such constitutive laws in their traditional physical statements; i.e., for the moment, we will assume that $T(M)$ has a Lorentzian structure defined by the tensor field $g$.

Since $\chi$ takes 2-forms to 2-vector fields and 2-forms can be evaluated on 2-vector fields, $\chi$ also defines a bilinear form on $\Lambda^2(M)$ by way of:

$$\chi(\alpha, \beta) = \alpha(\chi(\beta)). \qquad (4.6)$$

In local components, if $\alpha = 1/2\, \alpha_{\mu\nu} \theta^\mu \wedge \theta^\nu$ and $\beta = 1/2\, \beta_{\mu\nu} \theta^\mu \wedge \theta^\nu$ then this takes the form:

$$\chi(\alpha, \beta) = \chi^{\kappa\lambda\mu\nu} \alpha_{\lambda\mu} \beta_{\mu\nu} . \qquad (4.7)$$

On the surface of things, although physics usually assumes that $\chi$ is an invertible map, hence, the bilinear form $\chi$ is non-degenerate, nevertheless, it does not have to be symmetric. To see what this would mean physically, let us look at the electromagnetic field strength 2-form *F* in its *E-B* form for a given choice of unit future-pointing timelike vector field **t** and its metric dual 1-form θ:

$$F = \theta \wedge E - *(\theta \wedge B). \tag{4.8}$$

Of course, in this expression the * operator comes from the Lorentzian structure and the Lorentzian volume element:

$$\mathcal{V} = \sqrt{|-g|}\, dx^0 \wedge \ldots \wedge dx^3. \tag{4.9}$$

For now, we only consider the topologically trivial case where *M* is simply Minkowski space in order to concentrate on the application of our methodology developed above; the non-trivial case will then have to be treated in a subsequent study in which the emphasis will be on *projective differential* geometry.

If $\Pi_3$ is the spacelike 3-plane in $\mathbb{R}^4$ that is orthogonal to **t** then we have a decomposition of $\Lambda^2(\mathbb{R}^4) = \tilde{\Pi}_3 \oplus *\tilde{\Pi}_3 = \tilde{\Pi}_3 \oplus \Lambda^2(\Pi_3)$ into spacelike and timelike 2-forms, which can also be thought of as electric and magnetic or real and imaginary, respectively.

The 2-vector field that corresponds to *F* by way of the metric isomorphism of *T**(M*) with *T*(*M*) takes the form:

$$\mathbf{F} = \mathbf{t} \wedge \mathbf{E} - *(\mathbf{t} \wedge \mathbf{B}) = E_i \mathcal{E}^i - B_i *\mathcal{E}^i\,; \tag{4.10}$$

of course, the * operator in this case is the one that is defined by *g* and $\mathcal{V}$ on $\Lambda_2(\mathbb{R}^4)$, and their will be an analogous decomposition $\Lambda_2(\mathbb{R}^4) = \tilde{\Pi}_3 \oplus *\tilde{\Pi}_3 = \tilde{\Pi}_3 \oplus \Lambda_2(\Pi_3)$. The 2-forms $\mathcal{E}^i = \theta \wedge \theta^i$, $i = 1, 2, 3$ define a frame for $\tilde{\Pi}_3$, so the $*\mathcal{E}^i$ define a frame for $*\tilde{\Pi}_3$.

Furthermore, we decompose $\mathfrak{H} = \chi(F)$ as:

$$\mathfrak{H} = \mathbf{t} \wedge \mathbf{D} - *(\mathbf{t} \wedge \mathbf{H}). \tag{4.11}$$

With these decompositions of $\Lambda^2(\mathbb{R}^4)$ and $\Lambda_2(\mathbb{R}^4)$, one can express χ in block-matrix form ([5]):

$$[\chi] = \begin{bmatrix} \mathcal{A} & \mathcal{D} \\ \hline \mathcal{C} & \mathcal{B} \end{bmatrix}, \tag{4.12}$$

and the constitutive map takes the block-matrix form:

$$\begin{bmatrix} \mathbf{D} \\ \mathbf{H} \end{bmatrix} = \begin{bmatrix} \mathcal{A} & \mathcal{D} \\ \hline \mathcal{C} & \mathcal{B} \end{bmatrix} \begin{bmatrix} E \\ B \end{bmatrix}, \tag{4.13}$$

In the notation of Post, $\mathcal{A} = \varepsilon$ represents the electric permittivity matrix of the medium, $\mathcal{B} = \mu^{-1}$ is the inverse of the magnetic permeability matrix, and the off-diagonal blocks $\mathcal{C} = -\gamma = \mathcal{D}^T$ represent the coupling of the electric field to the magnetic induction and vice versa. Although this effect has not been observed in Nature for stationary media, nevertheless, it appears in the propagation of electromagnetic waves in moving media in the form of the Fresnel-Fizeau effect, the Faraday effects, and the existence of

---

[5] In order to minimize confusion in the literature, we use the notation of Hehl and Obukhov [**12**] for the block matrices.

natural optical activity. Hence, for stationary media one should expect a decoupling of $\chi$ into a direct sum of matrices and the constitutive equations would then take the form:
$$\mathbf{D} = \mathcal{A}(E), \qquad \mathbf{H} = \mathcal{B}(B). \tag{4.14}$$

In full generality, one should also approach the issue of symmetry by a decomposition of the tensor $c$ into a symmetric, "trace-free" part, $\chi^{(1)}$, an anti-symmetric part $\chi^{(2)}$ that Hehl and Obukhov refer to as a *skewon*, and a "trace" part $\chi^{(1)}$ that they identify as an *axion* field:
$$\chi = \chi^{(1)} + \chi^{(2)} + \chi^{(3)}, \tag{4.15}$$
where:
$$\chi^{(1)}(F, G) = \tfrac{1}{2}(\chi(F, G) + \chi(G, F)) - \chi^{(3)}(F, G), \tag{4.16a}$$
$$\chi^{(2)}(F, G) = \tfrac{1}{2}(\chi(F, G) - \chi(G, F)), \tag{4.16b}$$
$$\chi^{(3)}(F, G) = \alpha \mathcal{V}(F, G), \tag{4.16c}$$

in which we see the appearance of the scalar product $(F, G)$ in the axion contribution. This decomposition also gives us the irreducible representations of $GL(4;\mathbb{R})$ in $\Lambda^2(\mathbb{R}^4) \otimes \Lambda^2(\mathbb{R}^4)$. If one defines the trace of $\chi$ to be $\mathrm{Tr}(\chi) = \sum_{I=1}^{6} \chi(\varepsilon_I, \varepsilon_I)$ for any unit-volume frame $\varepsilon_I$, $I = 1, \ldots, 6$ that spans $\Lambda^2(\mathbb{R}^4)$ then one can also see that:
$$\alpha = \frac{1}{6}\mathrm{Tr}(\chi), \tag{4.17}$$
as intended.

We now see that a scalar product on $\Lambda^2(\mathbb{R}^4)$ can only involve the symmetric part of $\chi$, viz., $\chi^{(1)} + \chi^{(2)}$, which amounts to the vanishing of the skewon; there are also reasons to confine one's attention for just $\chi^{(1)}$. For stationary matter, the resulting scalar product decouples into a sum of "real" and "imaginary" scalar products under the decomposition of $\Lambda^2(\mathbb{R}^4)$ defined by $\mathbf{t}$ and $*$:
$$\chi^{(1)}(F, G) = \chi^{(1)}_{\mathrm{Re}}(\theta \wedge E, \theta \wedge E') - \chi^{(1)}_{\mathrm{Im}}(\theta \wedge B, \theta \wedge B'). \tag{4.18}$$

When the medium is, moreover, electrically and magnetically isotropic, this reduces to:
$$\chi^{(1)}(F, G) = \varepsilon(g \wedge g)(\theta \wedge E, \theta \wedge E') - \frac{1}{\mu}(g \wedge g)(\theta \wedge B, \theta \wedge B')$$
$$= \varepsilon g(E, E') - \frac{1}{\mu}g(B, B'). \tag{4.19}$$

Under an injection of $\mathbb{R}^{4*}$ into $\Lambda^2(\mathbb{R}^4)$ that is defined by the 1-form $\theta$, a complementary 3-plane $\Pi_3$ that contains $E$, $E'$, $B$, $B'$, and a choice of unit normal $n$ to $\Pi_3$, the scalar product on 2-forms that is defined by $\chi^{(1)}$ pulls back to a Lorentzian scalar product on $\mathbb{R}^{4*}$ of "electromagnetic" type:
$$\langle \alpha, \beta \rangle = -\frac{1}{\mu}\chi^{(1)}(\theta \wedge \alpha, \theta \wedge \beta) + \varepsilon \chi^{(1)}(\theta \wedge n, \theta \wedge n), \tag{4.20}$$

which is then conformal to the usual Lorentzian scalar product if one sets $c^2 = 1/\varepsilon\mu$.

   *c. Deriving the complex structure from the constitutive law.* For brevity, we now use the notation $\chi$ when we actually refer to its trace-free symmetric part $\chi^{(1)}$. The choice of a unit volume element $\mathcal{V}$ defines Poincaré duality #: $\Lambda_k(M) \to \Lambda^{*-k}(M)$, $\mathbf{a} \mapsto i_\mathbf{a}\mathcal{V}$. Although it is tempting to simply set $* = \# \circ \chi$, nevertheless, when we try this with a common choice of constitutive law, we see that when we evaluate $*^2 = \# \circ \chi \circ \# \circ \chi$ for the simplest constitutive law (4.19), we do not get $- I$, as desired:

$$\# \circ \chi \circ \# \circ \chi[E, B] =$$
$$\# \circ \chi \circ \#[\varepsilon\mathbf{E}, \frac{1}{\mu}\mathbf{B}] = \# \circ \chi[-\varepsilon B, \frac{1}{\mu}E] = \frac{\varepsilon}{\mu}\#[-\mathbf{B}, \mathbf{E}] = -\frac{\varepsilon}{\mu}[E, B].$$

Hence, in this case $*^2 = -\varepsilon/\mu\, I$; for more general $\chi$, we must replace $\varepsilon/\mu$ with a more general factor $\lambda^2$. Consequently, we must define the almost-complex structure as a scaled version of our tentative definition:

$$* = \lambda^{-1} \# \circ \chi. \tag{4.21}$$

   *d. The role of the scalar products on $\Lambda_2(\mathbb{R}^4)$ in electromagnetism.* All of the scalar products on $\Lambda_2(\mathbb{R}^4)$ that we defined above play a key role in electromagnetism, when one looks at the **D-H** form of $\mathfrak{H}$, relative to a choice of "unit timelike" vector field **t**:

$$\mathfrak{H} = \mathbf{t} \wedge \mathbf{D} - *(\mathbf{t} \wedge \mathbf{H}) = \mathfrak{H}_R - *\mathfrak{H}_I, \tag{4.22}$$

which corresponds to the element of $\mathbb{C}^3$:

$$\mathbf{G} = \mathbf{D} - i\mathbf{H}. \tag{4.23}$$

Of course, we need to clarify what we mean by "unit timelike" in the absence of a Minkowski scalar product, and one must first map **D** to $\mathbf{t} \wedge \mathbf{D}$ and **H** to $\mathbf{t} \wedge \mathbf{H}$. However, from our discussion above, we see that we are simply dealing with an injection of a 3-plane in $\mathbb{R}^4$ into $\Lambda_2(\mathbb{R}^4)$ of the form $e_\mathbf{t}$. Hence, we can define **t** to be timelike in terms of the scalar product (.,.) on $\Lambda_2(\mathbb{R}^4)$, and norm-squared of **t** plays no role in that process.

The quadratic forms that are associated with the inner products that we defined take the form:

$$\langle\mathfrak{H}, \mathfrak{H}\rangle = \mathbf{D}^2 - \mathbf{H}^2, \tag{4.24a}$$
$$(\mathfrak{H}, \mathfrak{H}) = 2\mathbf{D}\cdot\mathbf{H}, \tag{4.24b}$$
$$((\mathfrak{H}, \mathfrak{H})) = ((\mathfrak{H}, \mathfrak{H}))_\mathbb{C} = \mathbf{D}^2 + \mathbf{H}^2. \tag{4.24c}$$

The first is proportional to the electromagnetic field Lagrangian, the second gives the angle between **D** and **H** in the plane they span, and the last one is proportional to the energy density of the electromagnetic induction field.

We can use the complex Euclidian scalar product $\langle.,.\rangle_\mathbb{C}$ that we discussed in section **3**, part *d*, eq. (1.22) to combine the first two expressions in (4.24) into:

$$\langle\mathfrak{H}, \mathfrak{H}\rangle_\mathbb{C} = (\mathbf{D}^2 - \mathbf{H}^2) + 2i\, \mathbf{D}\cdot\mathbf{H}. \tag{4.25}$$

The 2-vectors that represent electromagnetic waves – i.e., the isotropic ones – are then characterized by being the ones that make this quadratic form vanish.

An intriguing aspect of the inner products that we constructed above is the fact the inner product on $\Lambda^2(\mathbb{R}^4)$ that gives us the electromagnetic field energy density also defines a Hermitian structure on the complex vector space associated with $\Lambda^2(\mathbb{R}^4)$. Since this also brings about a reduction from the action of $SL(3; \mathbb{C})$ on $\Lambda^2(\mathbb{R}^4)$ to an action of $SU(3)$, we should necessarily wonder if this fact is leading us from the realm of electromagnetic interactions into the realm of strong interactions, even though $SU(3)$ is usually totally foreign to any classical discussion of the electromagnetic interaction.

*e. The role of duality planes.* Now that we have established the physical relevance of the complex structure on $\Lambda^2(\mathbb{R}^4)$, and with it, the $\mathbb{C}$-linear isomorphism of $\Lambda^2(\mathbb{R}^4)$ with $\mathbb{C}^3$, we next discuss the physical nature of complex lines through the origin of $\Lambda^2(\mathbb{R}^4)$, i.e., duality planes.

If we use $F$ as a typical 2-form then the duality plane that is spans consists of all 2-forms of the form $\alpha F + \beta *F$, as $\alpha$ and $\beta$ range over all real scalars. We note that if we perform a linear transformation of $F$, $*F$ within this plane to:

$$F' = \alpha F + \beta *F \tag{4.26a}$$
$$*F' = -\beta F + \alpha *F, \tag{4.26b}$$

then if $F$, $*F$ satisfy the source-less Maxwell equations, so do $F'$, $*F'$. However, the matrix of this transformation takes the form:

$$\begin{bmatrix} \alpha & \beta \\ -\beta & \alpha \end{bmatrix} = \alpha I_2 + \beta J \tag{4.27}$$

that is characteristic of multiplication by a complex number. Hence, a source-less solution of the Maxwell equations is defined only up to a duality rotation – characterized by $\beta$ – and a real scalar multiplier $\alpha$.

In light of this ambiguity, it would be better to regard such solutions as equivalence classes of 2-forms under the action of complex scalars, i.e., complex lines through the origin of $\Lambda^2(\mathbb{R}^4)$. Hence, we see that the complex projectivization of $\Lambda^2(\mathbb{R}^4)$, which we have denoted by $\mathbb{C}P\Lambda^2(\mathbb{R}^4)$, represents the space of distinct equivalence classes of solutions, up to complex multiplication.

*f. Electromagnetic field types.* We have already observed that any non-zero bivector in $\mathbb{R}^4$ must have rank two or four. It is intuitively illuminating to see what sort of electromagnetic fields these types of bivectors can represent when one also chooses a 3+1 decomposition of $\mathbb{R}^4$ into $[\mathbf{t}] \oplus \Pi_3$. One finds that although the rank-four case is generic, in a sense, nevertheless, the rank-two case includes the more commonly considered elementary cases. We then say that (relative to the chosen decomposition) a given bivector $\mathbf{F}$ is:

   i. *Purely electric* iff $\mathbf{F} = \mathbf{t} \wedge \mathbf{E}$ for some $\mathbf{E} \in \Pi_3$.
   ii. *Purely magnetic* iff $\mathbf{F} = *_s \mathbf{B}$ for some $\mathbf{B} \in \Pi_3$.
   iii. *Isotropic* iff $\mathbf{F} = (\mathbf{t} + \mathbf{n}) \wedge \mathbf{E}$ for some $\mathbf{E}, \mathbf{n} \in \Pi_3$ and $<\mathbf{F}, \mathbf{F}>_\mathbb{C} = 0$.
   iv. *Generic* iff $\mathbf{F}$ is of rank four. In this case, it will take the form $\mathbf{F} = \mathbf{t} \wedge \mathbf{E} + *_s \mathbf{B}$.

The first three types are of rank two. Because of electromagnetic induction, the first two cases will be static in the rest system of **t**.  The isotropic case includes electromagnetic waves, but also the trivial cases of perpendicular static homogeneous electric and magnetic fields of equal magnitudes, as well as more unconventional cases in which the fields can vary, say, linearly in time.  Hence, the generic case must include both electric and magnetic fields that are either of unequal magnitudes, non-perpendicular, or not in a wavelike state.  The generic fields also define the six-dimensional complement to the Klein quadric in $\Lambda^2(\mathbb{R}^4)$, which then consists of two connected components, namely, $<\alpha, \alpha> < 0$ and $<\alpha, \alpha> > 0$.

*g. Electromagnetic energy-momentum tensor.* Although energy (at least kinetic energy, anyway) usually seems to suggest an unavoidable introduction of a metric structure, nevertheless, the (stress-)energy-momentum tensor for the electromagnetic field $F$ can be introduced and analyzed without the necessity of introducing such a device.  Rather than regarding an energy-momentum tensor as a possibly-degenerate, possibly-asymmetric scalar product or a map that takes the velocity vector field for some particular motion in spacetime to its corresponding energy-momentum covector, one simply regards it as a linear map from each tangent space to itself that takes tangent vectors, which represent either infinitesimal displacements or hypersurface element normals, to other tangent vectors, which then represent conserved currents or energy-momentum fluxes through the hypersurface element, respectively.  Actually, the units of the elements of the energy-momentum tensor can represent energy density, momentum flux, or pressure, depending on how one applies them.

The *Faraday energy-momentum tensor* that is associated with the electromagnetic field $F$, or rather with the electromagnetic field Lagrangian $F \wedge *F = <F, F>\mathcal{V}$, can be obtained from the set of four 3-forms:

$$\tau_\mu = i_{\mathbf{e}_\mu} F \wedge *F - F \wedge i_{\mathbf{e}_\mu} *F, \quad \mu = 0, 1, 2, 3, \tag{4.28}$$

in which we have made a choice of linear frame $\mathbf{e}_\mu$, $\mu = 0, 1, 2, 3$ for $\mathbb{R}^4$. However, from linearity, the expression is equivariant under the action of $GL(4; \mathbb{R})$, so the choice is relatively harmless.

Let us convert this expression slightly by introducing some useful 1-forms, namely:

$$f_\mu = i_{\mathbf{e}_\mu} F, \qquad \hat{f}_\mu = i_{\mathbf{e}_\mu} *F, \quad \mu = 0, 1, 2, 3. \tag{4.29}$$

One can then reconstruct $F$ and $*F$ from:

$$F = \tfrac{1}{2} \theta^\mu \wedge f_\mu, \qquad *F = \tfrac{1}{2} \theta^\mu \wedge \hat{f}_\mu, \tag{4.30}$$

in which $\theta^\mu$ is the coframe that is reciprocal to $\mathbf{e}_\mu$.

A particularly illuminating choice of frame is an *adapted frame*, for which $\mathbf{e}_0 = \mathbf{t}$ and $\mathbf{e}_i$ span $\Pi_3$.  One then sees that:

$$f_0 = E, \quad f_i = -E_i \theta^0 + \hat{B}_i, \qquad *_s B = \tfrac{1}{2} \theta^i \wedge \hat{B}_i, \tag{4.31a}$$

$$\hat{f}_0 = B, \quad \hat{f}_i = -B_i \theta^0 - \hat{E}_i, \qquad *_s E = \tfrac{1}{2} \theta^i \wedge \hat{E}_i, \tag{4.31b}$$

in which we have introduced the notations:
$$\hat{E}_i = i_{\mathbf{e}_i} *_s E, \qquad \hat{B}_i = i_{\mathbf{e}_i} *_s B. \qquad (4.32)$$

The 1-forms represent essentially projections of the spacelike 2-forms $*_s E$ and $*_s B$ into the directions $\mathbf{e}_i$.

Hence, one sees that one can interpret the two families of 1-forms, $f_\mu$ and $\hat{f}_\mu$, as essentially the field strengths of $F$ or $*F$ in the directions defined by the choice of frame. Similarly, since the Lorentz force associated with the interaction of $F$ with an electric current vector $\mathbf{J}$ is $i_\mathbf{J} F$, one could interpret the 1-forms as forces associated with the vector fields $\mathbf{e}_\mu$, when regarded as "unit currents;" of course, this is consistent with the notion that the field strengths are the forces that act one unit charges, whether electric or magnetic. Note that these four 1-forms will define a 4-coframe in their own right only if $F$ has rank four. In the rank two case, one can generally find a frame $\mathbf{e}_\mu$ for which the resulting coframes have two zero members.

In the general case, we can then rewrite the $\tau_\mu$ in the form:
$$\tau_\mu = \tfrac{1}{2}(f_\mu \wedge *F - F \wedge \hat{f}_\mu) = -\tfrac{1}{4}(f_\mu \wedge \hat{f}_\nu + f_\nu \wedge \hat{f}_\mu) \wedge \theta^\nu. \qquad (4.33)$$

Corresponding to the 3+1 decomposition of $F$ and $*F$ we then have a 3+1 decomposition of the $\tau_\mu$, namely:
$$\tau_0 = -*_s S \wedge \theta^0 + \tfrac{1}{2}(E^2 + B^2)\mathcal{V}_s \qquad (4.34a)$$
$$\tau_i = -\tfrac{1}{2}(E^2 + B^2) i_{\mathbf{e}_i} \mathcal{V}_s \wedge \theta^0 + i_{\mathbf{e}_i} *_s S. \qquad (4.34b)$$

We have introduced the *Poynting covector*:
$$S = *_s (E \wedge B) \qquad (4.35)$$

into these expressions. It is spacelike and represents the momentum flux carried by the field, although this interpretation is generally physically meaningful only in the case of electromagnetic waves.

When one applies Poincaré duality to the 3-forms $\tau_\mu$, one produces four vectors:
$$\mathbf{T}_\mu = \#\tau_\mu = T_\mu^\nu \mathbf{e}_\nu. \qquad (4.36)$$

In particular, for an adapted frame:
$$\mathbf{T}_0 = \tfrac{1}{2}(E^2 + B^2)\mathbf{t} - \mathbf{S} \qquad (4.37a)$$
$$\mathbf{T}_i = S_i \mathbf{t} - \{\tfrac{1}{2}(E^2 + B^2)\mathbf{e}_i - E_i \mathbf{E} - B_i \mathbf{B}\}. \qquad (4.37b)$$

Hence:
$$T_0^0 = \tfrac{1}{2}(E^2 + B^2), \qquad T_0^i = S^i, \qquad (4.38a)$$
$$T_i^0 = -S_i, \qquad T_i^j = -\tfrac{1}{2}(E^2 + B^2)\delta_i^j - E_i E^j - B_i B^j, \qquad (4.38b)$$

which agrees with the usual component expressions.

Note that we still have not needed to introduce a scalar product on $\mathbb{R}^4$, since the vectors $\mathbf{S}$, $\mathbf{E}$, and $\mathbf{B}$ were obtained from the 3-forms $*_s S \wedge \theta^0$ and $i_{\mathbf{e}_i} *_s S$ by Poincaré duality alone, not by using the scalar product that is induced on $\Pi_3$ by the injection that **t** defines.

If we regard the energy-momentum tensor in "mixed" form – i.e., the matrix $T_\mu^\nu$ – as the matrix of a linear endomorphism $\mathfrak{T}$ on $\mathbb{R}^4$ then we see that we have another way of getting around the full Minkowski structure on $\mathbb{R}^4$ by taking a "square root" of something more general (along with obtaining the Lorentzian structure on spacetime from an "exterior square root" of the constitutive map) is to assume that $\mathfrak{T}$ takes the conventional form:

$$\mathfrak{T} = \mathfrak{F}^2 - \tfrac{1}{2}<F, F> I, \tag{4.39}$$

for some linear endomorphism $\mathfrak{F}$. Hence, if $\mathfrak{F}$ exists then it will represent a square root of the operator:

$$\mathfrak{F}^2 = \mathfrak{T} + \tfrac{1}{2}<F, F> I. \tag{4.40}$$

The question of existence for an operator square root for $\mathfrak{T}$ reverts to the question of the existence of real eigenvalues. Since the matrix $\mathfrak{T}$ has the usual mixed form for the components of the Faraday tensor, one can borrow from the analysis of that object for further guidance.

The matrix $\mathfrak{T}$ is seen to have trace zero and satisfy the symmetry requirement that when one lowers one of its indices with $\eta_{\mu\nu}$, one produces a symmetric matrix. Hence, one can say that $\mathfrak{T}$ belongs to the vector subspace of the Lie algebra $\mathfrak{gl}(4)$ that consists of *infinitesimal* (*volume-preserving*) *Lorentz strains*. These are one of the two types of real 4×4 matrices that one produces by polarizing an arbitrary real 4×4 matrix $A$ by means of the *Lorentz adjoint* operator:

$$A^* = \eta A^T \eta, \tag{4.41}$$

the other being the infinitesimal Lorentz transformations. Hence, we see that the Lorentzian structure is already present to some degree.

The operator $\mathfrak{F}^2$ no longer has trace zero, but it still represents an infinitesimal Lorentz strain. If the operator $\mathfrak{F}$ takes the form of the usual matrix $F_\nu^\mu$ of mixed components for $F$ then it will be an element of $\mathfrak{so}(3,1)$, since when one lowers the upper index with $\eta_{\mu\nu}$, one will produce an anti-symmetric matrix. Hence, the association of $\mathfrak{F}$ with $F$ that we have defined is a roundabout way of raising one of the indices on $F$ without introducing a scalar product on $\mathbb{R}^4$, only on $\Pi_3$.

As for the issue of eigenvalues, the anti-symmetry of $F$ has the indirect consequence that the characteristic polynomial for F is:

$$\lambda^4 + 2<F, F> \lambda^2 + (F, F), \tag{4.42}$$

whose roots are obtained from:

$$\lambda^2 = -<F, F> \pm \sqrt{<F, F>^2 - (F, F)^2}. \tag{4.43}$$

If $F$ has rank two then $(F, F) = 0$ and the roots are either $\pm i<F, F>$ or 0, which then has multiplicity two. Clearly, if $F$ is isotropic then the only eigenvalue is zero.

*h. Electromagnetic waves* [**15**]. When one tries to make the Lorentzian metric of spacetime a consequence of the electromagnetic constitutive laws one finds that trying to discuss wavelike solutions to the Maxwell equations (4.5) is fraught with subtleties that only look like trivialities when one assumes that a Lorentzian structure is already present, or at least a conformal Lorentzian structure.

The first problem is encountered when one tries to derive the wave equation from Maxwell equations. If we substitute the constitutive law into the equation for $\mathfrak{H}$ then the source-free equations become:

$$dF = 0, \qquad d\#\chi(F) = 0 . \qquad (4.44)$$

From (4.20), this can be written:

$$dF = 0, \qquad d(\lambda^2 *(F)) = 0 . \qquad (4.45)$$

Even if we assume that $\lambda$ is constant, which is similar to, but not equivalent to, the assumption that $c_0$ is constant, so the second equation becomes simply the usual $d*F = 0$, we see that one cannot combine the two equations in the usual way to give $\Box F = \delta dF + d\delta F = 0$, since the definition of the codifferential operator $\delta = (-1)^{k(4-k)}*d*$ implies that the $*$ isomorphism must be defined on 1-forms and 3-forms, when we have only assumed that $*$, by way of $\chi$, is defined on 2-forms. This limitation also applies if one chooses to derive the electromagnetic equation for a choice of potential 1-form $A$, as well.

Clearly, it is sufficient to extend $*$ to an isomorphism of $\Lambda^1(\mathbb{R}^4)$ with $\Lambda^3(\mathbb{R}^4)$, and since # gives us an isomorphism of $\Lambda_1(\mathbb{R}^4)$ with $\Lambda^3(\mathbb{R}^4)$, this means we need only define an isomorphism of $\Lambda_1(\mathbb{R}^4)$ with $\Lambda^1(\mathbb{R}^4)$, i.e., $\mathbb{R}^4$ with $\mathbb{R}^{4*}$. However, the symmetric case is essentially the same thing as introducing a Lorentzian structure, at least up to a conformal factor. Hence, in order for the d'Alembertian operator to be well-defined it is necessary and sufficient for the spacetime manifold ($\mathbb{R}^4$, in the present case) to have a conformal Lorentzian structure. Of course, this might seem tautological if one considers that the existence of electromagnetic waves implies the existence of light cones.

One possible way around the aforementioned dilemma is to consider first-order wave equations, such as conservation laws often define. For instance, if one factors the two-dimensional d'Alembertian in the obvious way then the two-dimensional wave equation:

$$u_{tt} - u_{xx} = 0 \qquad (4.46)$$

gives way to two first-order partial differential equations of the form:

$$u_t \pm u_x = 0 . \qquad (4.47)$$

The solutions to these equations are the individual traveling waves that add together to give the general solution of (4.46). Hence, the real essence of the traveling wave solutions is attributable to first-order differential equations, not second-order ones ([6]).

---

[6] Of course, one can also convert a second order differential equation into a system of first order equations by the introduction of derivative variables, which amounts to defining a new wavefunction on the first-order jet space of the original one.

Another way that one can reduce the order from second to first is actually equivalent to the Dirac factorization of the d'Alembertian into the square of the massless Dirac operator. One simply notes that the usual d'Alembertian can be factored into:

$$\Box = (d + \delta)^2 \equiv \slashed{d}^2 . \qquad (4.48)$$

This $\slashed{d}$ operator we have defined may not look like the Dirac operator $\gamma^\mu \partial_\mu$ on the surface of things – at least to mainstream physicists – but when one examines the linear isomorphism of the exterior algebra $\Lambda^*(\mathbb{R}^4)$ with the Clifford algebra over Minkowski space, which can be represented by the algebra of 4×4 complex matrices using the four $\gamma^\mu$ matrices as generators (cf. Benn and Tucker [**16**]), one finds that this is indeed the case.

Clearly, the massless Dirac equation that one obtains for $F$ from $\slashed{d}$, namely ([7]):

$$\slashed{d} F = dF + \delta F = 0 , \qquad (4.49)$$

is equivalent to the source-free Maxwell equations. In the present case, in which we can define $d^*$, but not $\delta$, we are dealing with an operator of the form $d + d^*$, whose square – namely, $d^*d + d^*d^*$ – can be defined if and only if $\delta$ is defined.

Another way of representing the same situation is to use the self-adjoint (or even anti-self-adjoint) part of the complexification of $F$, namely, $F_\pm = \frac{1}{2}(F \mp i*F)$, in which case, the Maxwell equations are equivalent to:

$$dF_\pm = 0 . \qquad (4.50)$$

Note that if we did not include the $i$ in the expressions for $F_\pm$ this equivalence would break down.

A necessary, but not sufficient, condition for a solution $F$ of Maxwell's equations to represent an electromagnetic wave is that $F$ be isotropic; i.e.:

$$\begin{aligned} 0 = <F, F>_\mathbb{C} &= <F, F> + i(F, F) \\ &= <F_{\text{Re}}, F_{\text{Re}}> - <F_{\text{Im}}, F_{\text{Im}}> + 2i <F_{\text{Re}}, F_{\text{Im}}>. \end{aligned} \qquad (4.51)$$

Hence, $F$ must be of rank two, i.e., decomposable, and its real and imaginary parts must have the same norm relative to the scalar product on $\Lambda^2(\mathbb{R}^4)$ that is defined by $\alpha \wedge *\beta$. The reason that this is not sufficient is because one can define such arrangements as constant electric and magnetic fields that have the same norm and are perpendicular, and trivially solve the Maxwell equations. The necessity follows from the fact that in order for $F$ to define a wavelike motion, it must define a lightlike propagation covector $l = \theta + n$ in some canonical manner. We have seen that is $F$ is isotropic then it can be given the form $l \wedge E$ for an appropriate 1-form $E$:

$$F = l \wedge E = \theta \wedge E + n \wedge E . \qquad (4.52)$$

Hence, we have:

$$F_{\text{Re}} = \theta \wedge E , \qquad F_{\text{Im}} = n \wedge E = *_s B. \qquad (4.53)$$

---

[7] In order to make sense of the resulting expression $dF + \delta F$, one simply regards it as a mixed exterior form in the exterior algebra over $\mathbb{R}^4$, as opposed to a homogeneous one. If one subdivides $\Lambda^*(\mathbb{R}^4)$ into even and odd-degree exterior forms, they will correspond to the even and odd elements in the Clifford algebra of Minkowski space.

When one chooses this time-orientation 1-form $\theta$, moreover, i.e., a class of rest frames for some physically-defined motion, the annihilating subspaces of $l$ and $\theta$ intersect in a 2-plane in $\mathbb{R}^4$ that one calls the *plane of polarization* for the wave. It will be the plane spanned by $E$ and $B$ and will be tangent to the spatial isophase surfaces ([8]), as well as the unit sphere in $\Pi_3 = \text{span}\{E, B, n\}$ at the point described by $n$.

Clearly, a wavelike $F$ must belong to the Klein quadric, as well as the complex quadric defined by $\langle .,. \rangle_{\mathbb{C}}$.

A mathematical device that physicists use in the discussion of wavelike solutions to the Maxwell equations, especially plane-wave solutions is the introduction of complex field strengths for $E$ and $B$. As it is usually discussed, this seems to be largely a matter of mathematical convenience more than a matter of deep physical principles that would necessitate this step, but we see that when one uses the complex vector space defined by $\Lambda^2(\mathbb{R}^4)$ and $*$ as an isomorphic copy of $\mathbb{C}^3$, one can give this complexification a somewhat more natural setting.

For instance, instead of setting $\mathcal{E} = E_{\text{Re}} + iE_{\text{Im}}$, as is usually practiced [**15**], One could represent $\mathcal{E}$ by the 2-form $\theta \wedge E_{\text{Re}} + *(\theta \wedge E_{\text{Im}})$. Although the imaginary part would seem to present a "magnetic" aspect to it, one could argue that the usual prescription that "the physical part of $\mathcal{E}$ is the real part" takes the form of saying that the measurable part of $\mathcal{E}$ is the projection of $\mathcal{E}$ into the spacelike 3-plane $\Pi_3$ of the 1-forms $E_{\text{Re}}$ and $E_{\text{Im}}$ by way of the composition of the projection of $\mathcal{E}$ onto $\theta \wedge E_{\text{Re}}$ and the inverse of the injection that takes $\Pi_3$ to $\theta \wedge \Pi_3$.

One then sees that the action of the phase rotation group $e^{i\varphi}$ on $\mathcal{E}$ is simply a rotation in the duality plane of $\mathcal{E}$; i.e., the 2-plane in $\Lambda^2(\mathbb{R}^4)$ that is spanned by the 2-forms $\theta \wedge E_{\text{Re}}$ and $*(\theta \wedge E_{\text{Im}})$.

*i. Fresnel quartics* [**12, 18, 19**]. There are actually good reasons to believe that the fundamental equations of electromagnetism might be *fourth* order partial differential equations, not second order ones. Indeed, this situation is somewhat reminiscent of the way that the fundamental equations of elastostatics start out as fourth order equations, such as the biharmonic equation $\Delta^2 \phi = 0$, and only factor into second order equations by making simplifying assumptions about the nature of the mechanical constitutive law of the elastic medium in question. Although one immediately thinks of causality concerns when discussing the order of differential equations in physics, nevertheless, one must also remember that it is the propagation of electromagnetic waves that seems to define causality to begin with.

In the present context, we are concerned with the electromagnetic constitutive law of the electromagnetic vacuum state, which differs from the mechanical law by the fact that stress and strain tensors are generally assumed to be symmetric second-rank tensors (in the absence of external torques on the medium), whereas the induction and field strength tensors of electromagnetism are assumed to be anti-symmetric. In either case, in the linear regime one is dealing with a four-rank tensor that represents a scalar product on a space of second-rank tensors and gives way to a fourth order partial differential equation whose symbol is (indirectly) defined by the constitutive law.

---

[8] For a more detailed discussion of wave structures on manifolds, cf. Delphenich [**17**].

One way to arrive at this conclusion is to use the Hadamard program for treating waves as perturbations in a medium that are characterized by the appearance of a discontinuity in some order of derivative. One deals with *jump* discontinuities, in which the value of $f_+(x) - f_-(x)$ is finite for all points $x \in \Sigma$, where we have denoted the left and right limits in the function $f$ by $f_+(x)$ and $f_-(x)$, and $\Sigma$ is a hypersurface in our manifold in question. One also assumes that $\Sigma$ is orientable, so one can characterize its tangent spaces by the annihilating subspaces of some non-zero 1-form $k$, which plays the role of the propagation 1-form for the wave front that is described by the hypersurface $\Sigma$.

In the case of electromagnetism with inductions, the Hadamard conditions on the fields $F$ and $H$ are simply that they are continuous across $\Sigma$, but suffer a jump discontinuity in their first derivatives:

$$[F] = 0, \quad [dF] = k \wedge f, \tag{4.54a}$$
$$[H] = 0, \quad [dH] = k \wedge h, \tag{4.54b}$$

for some 2-forms on $\Sigma$ that we denote by $f$ and $h$.

We pause to note that we have actually introduced a sort of analogue of the Fourier transform in the form the discontinuity bracket on $\Sigma$. That is, although the bracket does not turn $F$ ($H$, resp.) into $f$ ($h$, resp.), nevertheless, the bracket seems to turn exterior derivatives into exterior products if one regards the 2-form $f$ ($h$, resp.) as the "transform" of $F$ ($H$, resp.). Indeed, as we shall see, the analogy continues.

If one takes into account the vacuum Maxwell equations, in the form, $dF = dH = 0$, then one immediately derives the consequence that:

$$k \wedge f = 0, \quad k \wedge h = 0, \tag{4.55}$$

which also have the appearance of the Fourier-transformed Maxwell equations.

The latter equations are equivalent to the equations:

$$f = k \wedge a, \quad h = k \wedge b, \tag{4.56}$$

for appropriate 1-forms $a$ and $b$ that defined only on $\Sigma$. Not only do these equations resemble the Fourier transforms of the differential equations that express $F$ and $H$ as exterior derivatives of potential 1-forms $A$ and $B$, but one sees that in (4.56) the 1-forms $a$ and $b$ are not unique, but can be replaced by "gauge" transformed 1-forms of the form $a + \lambda k$, $b + \nu k$, expressions that are consistent with the gauge transformations of $A$ and $B$ into $A + d\lambda$ and $B + d\nu$, respectively.

If we now insert the constitutive law $H = \chi(F)$ into the Maxwell equations, the Hadamard condition for $H$ becomes:

$$[dH] = k \wedge \chi(f) = k \wedge h. \tag{4.57}$$

Hence, we have that $h = \chi(f)$, up to algebraic gauge transformations.

If we substitute the expression for $f$ from (4.56), we then get the new algebraic form for Maxwell's equation for $H$:

$$k \wedge \chi(k \wedge a) = 0. \tag{4.58}$$

This algebraic equation for $a$ can be solved only if $k$ satisfies the *Fresnel equation:*

$$\mathcal{G}(\chi)(k, k, k, k) = 0, \tag{4.59}$$

in which $\mathcal{G}$ is a fourth degree tensor density of weight 1 that one calls the *Tamm-Rubilar tensor density*. We shall not elaborate on its specific expression, but refer the reader to Hehl and Obukhov [**12**], and simply characterize it by the properties that it is completely symmetric and ultimately depends upon the electromagnetic constitutive law.

Since $\mathcal{G}$ is completely symmetric, one can regard (4.59) as a quartic polynomial equation in the four variables $k_\mu$; hence, it defines a quartic hypersurface in $\mathbb{R}^4$. By the homogeneity of this polynomial, we also define a quartic hypersurface in $\mathbb{RP}^3$, although the polynomial in the inhomogeneous coordinates $X_i = k_i/k_0$ is not generally homogeneous.

If we pursue our Fourier analogy, we see that $\mathcal{G}$ also becomes the symbol of the fourth order linear differential operator:

$$\mathcal{G}_{\mu\nu\rho\sigma} \frac{\partial}{\partial x^\mu} \frac{\partial}{\partial x^\nu} \frac{\partial}{\partial x^\rho} \frac{\partial}{\partial x^\sigma}. \tag{4.60}$$

Just as we wished to recover the Lorentzian structure of spacetime above by factoring * into essentially an exterior product of $g \wedge g$, we now wish to examine the possibility that the operator (4.60) might factor into a product of wave operators with possibly differing metrics:

$$\mathcal{G}^{\mu\nu\rho\sigma} \frac{\partial}{\partial x^\mu} \frac{\partial}{\partial x^\nu} \frac{\partial}{\partial x^\rho} \frac{\partial}{\partial x^\sigma} = \left( g_1^{\mu\nu} \frac{\partial}{\partial x^\mu} \frac{\partial}{\partial x^\nu} \right) \left( g_2^{\rho\sigma} \frac{\partial}{\partial x^\rho} \frac{\partial}{\partial x^\sigma} \right), \tag{4.61}$$

as well as the possibility that the two resulting metrics might coincide:

$$\mathcal{G}^{\mu\nu\rho\sigma} \frac{\partial}{\partial x^\mu} \frac{\partial}{\partial x^\nu} \frac{\partial}{\partial x^\rho} \frac{\partial}{\partial x^\sigma} = \left( g^{\mu\nu} \frac{\partial}{\partial x^\mu} \frac{\partial}{\partial x^\nu} \right) \left( g^{\rho\sigma} \frac{\partial}{\partial x^\rho} \frac{\partial}{\partial x^\sigma} \right). \tag{4.62}$$

The former possibility is referred to variously as *birefringence* or *bimetricity*. The difference between the two terms amounts to the difference between the optical phenomenon that appears in some crystals, for which the index of refraction – i.e., the speed of propagation – of the electromagnetic wave depends on its polarization state and the mathematical phenomenon of the association of two light cones to each point.

The real issue is part physics and part algebra. The physical part is taken up with the fact that the form of $\mathcal{G}$ is going to depend upon the electromagnetic constitutive law of the medium in question. Furthermore, only the "traceless" part $\chi^{(1)} + \chi^{(2)}$ of $\chi$ contributes to $\mathcal{G}$, but not the axion term $\chi^{(3)}$.

The algebraic part is due to the fact that we are really concerned with the reducibility of homogeneous quartic polynomials, and, in particular, whether they will factor into a product of distinct or identical quadratic polynomials of Lorentzian form.

Some progress has been made along these lines by Hehl, Obukhov, and Rubilar. For instance, Obukhov and Rubilar [**18**] showed that any $\mathcal{G}$ that is obtained from a nonlinear electromagnetic Lagrangian of the form $\mathscr{L}(I_1, I_2)$, where $I_1 = F \wedge F$, $I_2 = F \wedge *F$ will be reducible to a product of quadratic tensors of Lorentzian form; i.e., all such nonlinear models exhibit birefringence. The Born-Infeld theory is of the form in question, namely:

$$\mathscr{L}_{BI} = E_c^2 \left( \sqrt{1 + \frac{1}{2E_c^2} I_1 - \frac{1}{16E_c^4} I_2^2} - 1 \right), \tag{4.63}$$

and one finds that the birefringence is degenerate in this case; i.e., the two Lorentzian metrics that result from factorization coincide.

Obukhov and Rubilar also show that this is true for the nonlinear Lagrangians in question iff the linear map $\kappa = \#\chi$ that takes 2-forms on the discontinuity hypersurface $\Sigma$ to 2-forms on $\Sigma$ and is defined by the traceless part of the jump tensor $\chi$ satisfies the *closure property*: $\kappa^2 = \lambda I$.

*j. Effective metrics for nonlinear photons.* Some interesting results concerning the propagation of non-linear electromagnetic waves that are related to the foregoing ones were obtained by Novello and Salim [**20, 21**]. The basic idea is that not only are such waves subject to the constraints on lightlike propagation that is imposed by the background gravitationally-defined conformal Lorentzian structure, but when the field strengths are strong enough for vacuum polarization to occur, this conformal Lorentzian structure is altered by a local "effective" metric that comes about as a result of the vacuum polarization. Hence, one has a manifestation of nonlinearity in which the electromagnetic field changes the conformal structure that determines the form of its propagation; hence, the nonlinearity is essentially a self-interaction contribution.

The method of derivation for this result started out in the same place as in the previous section. The point of divergence was in the 3+1 decomposition of $F$ and $\mathcal{H}$ and choice of a specific nonlinear constitutive law. Hence, for a timelike unit four-velocity **v**, whose covelocity $\theta = i_t \eta$ is obtained from the background metric $\eta$, one decomposes $F$ and $\mathcal{H}$ according to:

$$F = \theta \wedge E + {}^*(\theta \wedge B), \qquad \mathcal{H} = \theta \wedge D + {}^*(\theta \wedge H), \tag{4.64}$$

and assumes the isotropic, but nonlinear, constitutive law:

$$D = \varepsilon(E)E, \qquad H = \mu^{-1}B, \tag{4.65}$$

in which the electric permittivity $\varepsilon$ depends upon the electric field strength, but the magnetic permeability $\mu$ is a constant.

The Hadamard analysis gives us that the characteristic equation for such a field that obeys Maxwell's equations is:

$$g(k, k) = 0, \tag{4.66}$$

in which one introduces an effective metric $g$ that is defined by:

$$g = \eta - (1 - (1+\xi)\mu\varepsilon)\mathbf{v} \otimes \mathbf{v} - \xi \mathbf{\varepsilon}_1 \otimes \mathbf{\varepsilon}_1, \tag{4.67}$$

where $\varepsilon' \equiv d\varepsilon/dE$, $x = \varepsilon' \|E\|/\varepsilon$, and $\mathbf{\varepsilon}_1$ is the unit vector that defines the direction in the polarization plane along which **E** varies.

The speed of propagation for such a nonlinear electromagnetic wave becomes:

$$c' = c\sqrt{\frac{1+\xi\cos^2\alpha}{1+\xi}}, \tag{4.68}$$

in which $c^2 = 1/\mu\varepsilon$ and $\alpha$ is the angle between **E** and **k**, which is the spacelike part of the propagation vector, i.e., the normal to the spatial wavefront.

In the limiting case for which $\varepsilon$ is constant – i.e., a linear isotropic homogeneous constitutive law – the effective metric $g$ reduces to the form:

$$g = \eta - (1 - \mu\varepsilon)\mathbf{v} \otimes \mathbf{v}, \tag{4.69}$$

that was obtained by Gordon [**22**] for the propagation of light waves in moving linear dielectrics.

**4. Discussion.** The conclusion of the last section brings us to the most conspicuous direction for expansion of the methodology proposed in this article, at least as far as the physics of electromagnetism is concerned: the extension of the methodology to nonlinear electromagnetic constitutive laws. Such a constitutive law would generally take the form of a diffeomorphism of $\Lambda_2(\mathbb{R}^4)$ with itself, although that would imply that the differential map would represent a linear constitutive law for small perturbations of the bivectors. Presumably, one would want to restrict oneself to diffeomorphisms that produce differential maps that behave like the linear electromagnetic constitutive laws that were treated here. In particular, they should each differ from a complex structure by a scalar factor. By the functoriality of differentiation, this suggests that such a diffeomorphism $\Xi$ must satisfy:

$$\Xi^2 = -\lambda I, \tag{5.1}$$

in which $\lambda$ is a positive scalar function.

Since nonlinearity is a vast and ill-defined category in pure mathematics, it would probably be advisable to start with the simplest forms of nonlinearity that seem to relate to electromagnetism, such as the constitutive law used in the previous discussion or the Born-Infeld case. It might be useful to model the creation/annihilation of electron-positron pairs as a phase transition that produces electric (and possibly magnetic) dipoles that would account for the nonlinearity in the vacuum constitutive laws. A reasonable direction to follow in this regard would be the various effective Lagrangians that one obtains from quantum electrodynamics, such as the Euler-Heisenberg Lagrangian that represents the quantum analogue of the Born-Infeld Lagrangian.

Something else that is unavoidable in any discussion of electromagnetism is the role of the spinorial representations that seem to be unavoidable, at least for the description of the fields of charged regions of spacetime. Some of the links between projective geometry and spinors, or Clifford algebras, in general, have been indicated above, and once again one sees that they are closely related to the study of isotropic 2-forms and conformal geometry, but more details of the relationship with projective geometry need to be specified. Some work along these lines had been done earlier by Veblen and Hlavaty.

Finally, one of the most intriguing aspects of the foregoing analysis is that one finds the group $SU(3)$ emerging in an otherwise classical, but quite natural, fashion, namely, it is the group of linear transformations of $\Lambda_2(\mathbb{R}^4)$ that preserve a given complex structure, as well as the Hermitian form, which represents the energy density of the electromagnetic field – i.e., its Hamiltonian. Since $SU(3)$ is usually first introduced in the context of quantum chromodynamics to account for the color gauge of the strong interaction, it seems surprising that one can introduce at a much earlier stage than that. Perhaps there is a path to electro-strong unification to be derived from this. This seems reasonable when one considers that one of the first points in history when the notion of a strong interaction of limited spatial range might be necessary was in the early days of atomic theory when it was realized that a nucleus of charged protons would not be stable without some other force to counteract the electrostatic repulsion. In fact, this argument is closely analogous to the one that was proposed by Poincaré in order to stabilize the charge distribution for an extended electron. The possible conjecture is that linear electromagnetism must be "completed" in the regime of small neighborhoods of elementary charge distributions by a nonlinear contribution that counteracts Coulomb repulsion over some small neighborhood. One might then suspect that the strong interaction amounts to the nonlinear part of the electromagnetic interaction. Undoubtedly, the appearance of *SU*(3) in the context of "classical" electromagnetism, combined with more knowledge of nonlinear electromagnetic constitutive laws might lead in that direction.

The next step in this study of the pre-metric geometry of spacetime and its relationship to electromagnetism is to examine the role of projective affine geometry in pre-metric electromagnetism. At a number of points in the foregoing discussions, allusions were made to the fact that projective geometric notions seem to play a fundamental role. In the next installment of this ongoing research, these notions will be introduced at the level of special relativity and then, just as projective geometry resolves to affine, conformal, and metric geometry as special cases, the projective geometric notions will be related to affine, conformal, and metric relativity, as well as electromagnetism. Of particular interest is the application of notions from complex projective geometry to the geometry of the space of bivectors on $\mathbb{R}^4$ when it is given a complex structure.

Finally, one must confront the inevitable transition from projective affine geometry to projective differential geometry. This phase of the research is not as straightforward as it sounds, since one must realize that the very definition of a differential structure on a differentiable manifold was only intended to generalize affine geometry, not projective geometry. Hence, one must consider one's very starting point more shrewdly that one might expect. The results of this study will follow as an eventual sequel to the aforementioned work on projective affine geometry and electromagnetism.

# References

1. Cartan, E., *On manifolds with an affine connection and the theory of relativity,* (English translation by A. Ashtekar of a series of French articles from 1923 to 1926) Bibliopolis, Napoli, 1986.
2. Van Dantzig, D., "The fundamental equations of electromagnetism, independent of metrical geometry," Proc. Camb. Phil. Soc. **30** (1934), 421-427.
3. Kottler, F. "Maxwell'sche Gleichungen und Metrik," Sitz. Akad. Wien IIa, **131** (1922), 119-146.
4. Debever, R., "Le rayonnement gravitationnel," Cahier de Physique, t. **18** (1964), 303-349.
5. Debever, R., "A Complex Vectorial Formalism in General Relativity," J. Math. Mech. **16** (1967), 761-785.
6. Israel, W. *Differential Forms in General Relativity,* Dublin Institute for Advanced Studies, 1970.
7. Penrose, R., Rindler, W., *Spinors and Spacetime, v. 1: two-spinor calculus and relativistic fields,* Cambridge University Press, Cambridge, 1984.
8. Ward, R.S., Wells, R.O. *Twistor Geometry and Field Theory,* Cambridge Univ. Press, Cambridge, 1990.
9. Harnett, G., "Metrics and Dual Operators," J. Math. Phys. **32** (1991), 84-91.
10. Lichnerowicz, A., *Théorie relativiste de la gravitational et de l'electromagnetisme,* Masson and Co., Paris, 1955.
11. Lichnerowicz, A,. "Ondes et radiations électromagnétiques et gravitationelles en relativité génerale," Annali di Matematica **50** (1960), 1-95.
12. Hehl, F., Obukhov, Y., *Foundations of Classical Electrodynamics,* Birkhäuser, Boston, 2003.
13. Delphenich, D.H., "On the Axioms of Topological Electromagnetism," LANL Archive, hep-th/0311256.
14. Post. E. J., *Formal Structure of Electromagnetics*, Dover, NY, 1997.
15. Jackson, J.D., *Classical Electrodynamics, 2$^{nd}$ ed.,* Wiley, NY, 1976.
16. Benn, I.M., Tucker, R.W., *An Introduction to Spinors and Geometry, with Applications to Physics,* Adam Hilger, Bristol, 1987.
17. Delphenich, D.H., "Foliated Cobordism and Motion," LANL Archive, gr-qc/0209091.
18. Obukhov, Y.N., Rubilar, G.F., "Fresnel Analysis of the wave propagation in nonlinear electrodynamics," LANL Archives, gr-qc/0204028.
19. Visser, M., Barcelo, C., Liberati, S., "Bi-refringence versus bi-metricity," LANL Archive gr-qc/0204017.
20. Novello, M., Salim, J.M., "Effective electromagnetic geometry," Phys. Rev. D, **63** (2001) 083511.
21. De Lorenci, V.A., Klippert, R., Novello, M., Salim, J.M., "Light propagation in nonlinear electrodynamics," LANL Archive, gr-qc/0005049.
22. Gordon, W., Ann. Phys. (Leipzig) **72** (1923), 421.